\documentclass{article}
\usepackage[top=2cm,bottom=2.5cm,right=3cm,left=3cm]{geometry}
\usepackage{graphicx}
\usepackage{multirow}
\usepackage{subfig}
\usepackage{psfrag}
\usepackage{rotating}
\usepackage{amsmath}
\usepackage{algpseudocode}
\usepackage{tkz-graph}
\usetikzlibrary{arrows}
\usepackage{xcolor,colortbl}
\usepackage[T1]{fontenc}
\usepackage[T1]{fontenc}
\usepackage[utf8]{inputenc}
\usepackage{natbib}
\usepackage{pifont} 
\usepackage{amsmath,amssymb}
\usepackage{authblk}
\usepackage{natbib}

\numberwithin{equation}{section}

\title{Modelling the clustering of extreme events for short-term risk assessment}
\author[1]{Ross Towe\thanks{r.towe@lancaster.ac.uk}}
\author[2]{Jonathan Tawn}
\author[2]{Emma Eastoe}
\author[3,4]{Rob Lamb}
\affil[1]{School of Computing and Communications, Lancaster University, United Kingdom}
\affil[2]{Department of Mathematics and Statistics, Lancaster University, United Kingdom}
\affil[3]{JBA Trust, Skipton, United Kingdom}
\affil[4]{Lancaster Environment Centre, Lancaster University, United Kingdom}
\begin{document}
\maketitle
\begin{abstract}
Having reliable estimates of the occurrence rates of extreme events is highly important for insurance companies, government agencies and the general public. The rarity of an extreme event is typically expressed through its return period, i.e., the expected waiting time between events of the observed size if the extreme events of the processes are independent and identically distributed. A major limitation with this measure is when an unexpectedly high number of events occur within the next few months immediately after a \textit{T} year event, with \textit{T} large. Such events undermine the trust in the quality of these risk estimates. The clustering of apparently independent extreme events can occur as a result of local non-stationarity of the process, which can be explained by covariates or random effects. We show how accounting for these covariates and random effects provides more accurate estimates of return levels and aids short-term risk assessment through the use of a new risk measure, which provides evidence of risk which is complementary to the return period. 
\end{abstract}


\section{Introduction}\label{sec:Intro}
Floods, and other extreme weather-related hazards are often described in terms of their return period; i.e., the expected waiting time between events if the processes being described are assumed to be stationary. In recent years, the annual exceedance probability (AEP) has also been used as an alternative metric to communicate the severity of an extreme event, because it emphasises that the risk is present in any year. When the process is changing over time, such as with a trend, e.g., due to climate change, the return period is not a particularly helpful measure of risk, but the AEP is still useful, it simply changes from year to year. When the process has such non-stationarity a number of alternative ways of best presenting the associated risk over time have been proposed \citep{Rootzen2013, Cheng2014}. Here, we examine deficiencies in the return period and AEP descriptions for stationary process when separate independent flood events cluster in time, and we propose an additional risk measure to address the consequent difficulty in properly communicating this risk.

During the winter of 2015/2016, the north west of the United Kingdom experienced a sequence of storm events known as Desmond, Eva and Frank. Sustained heavy rainfall caused approximately $\pounds1.6$billion of economic damages, flooded 21,000 properties and severed major transport routes including the west coast train line being closed for a number of months \citep{EA2018}. For some communities, it was the latest in a series of repeated flood events, such as in Carlisle, which had flooded badly in 1979, 1980, 1984 and 2005. Similar experiences have been felt elsewhere in the UK e.g., for Harbertonford on the river Harbourne in Devon, a number of Environment Agency studies have tried to determine how best to alleviate the risk of flooding. This is because Harbertonford was flooded 21 times in the last 60 years and from 1998 to 2000 was flooded on six individual occasions \citep{Bradley2005,EA2012}.

This reoccurrence of extreme events illustrates long-term problems with the miscommunication of risk to the general public and decision makers \citep{Vogel2011,Cooley2013,Oleson2015}, which stems from a historical confusion of long- and short-term risks through a single risk measure, namely the return period. The return level $z_{T}$ is associated with the return period $T$, through the distribution function $G$ of the annual maximum by $G(z_{T})=1-1/T$ and the AEP being $T^{-1}$. For this paper we refer to $T$ as being the return period. Strictly $T$ is the return period of the annual maxima, with $T_{ALL}=-1/\log\left(1-T\right)$ being the expected waiting time  between exceedances of $z_{T}$ for all of the data, where $T_{ALL}\approx T-0.5$ for $T>20$. A major limitation with the return period as a risk measure is that there are regular occurrences when within the few months following a $T$ year event ($T > 50$ year) another event of similar or greater severity occurs. This undermines the reputation of statisticians and flood risk managers. 

The problem is that the current method of communicating the severity of an event focusses on the process being stationary, whereas there are many examples where this is not the case \citep{Black1997,Hannaford2012,Gilleland2017}. Clustering of apparently independent extreme events exists as a result of local non-stationarity of the process. Flood risk managers need methods to assess the short-term risk of the reoccurrence of flooding after large events have been observed, which the return period is unable to capture if $G$ changes from year-to-year. 

In some environmental applications the standard convention is to model the maxima of the data as being identically distributed and to ignore the potential effects of covariates. There are two different univariate approaches: either a block maxima based approach or the method considered here, which looks at exceedances above a predetermined threshold. We adopt the latter approach due to the efficiency gains in maximising the available information in the data as well as it providing an analysis at a temporal scale for which detecting covariate-response relationships is easier \citep{Coles2001}. There are two alternative threshold approaches, the generalised Pareto distribution \citep{Davison1990} and the non-homogeneous Poisson process \citep{Smith1989}. We adopt the latter approach as it is most easy to parametrise non-stationarity. For example, a linear trend is accounted for in the location parameter of the Poisson process approach unlike for the generalised Pareto approach which instead needs both the rate of exceedance and the scale parameters to change in a non-linear fashion \citep{Coles2001}. 

\begin{figure}[!htbp]
\centering
\subfloat[][]{\label{fig:ClExt}
\includegraphics[width=8cm,height=5cm]{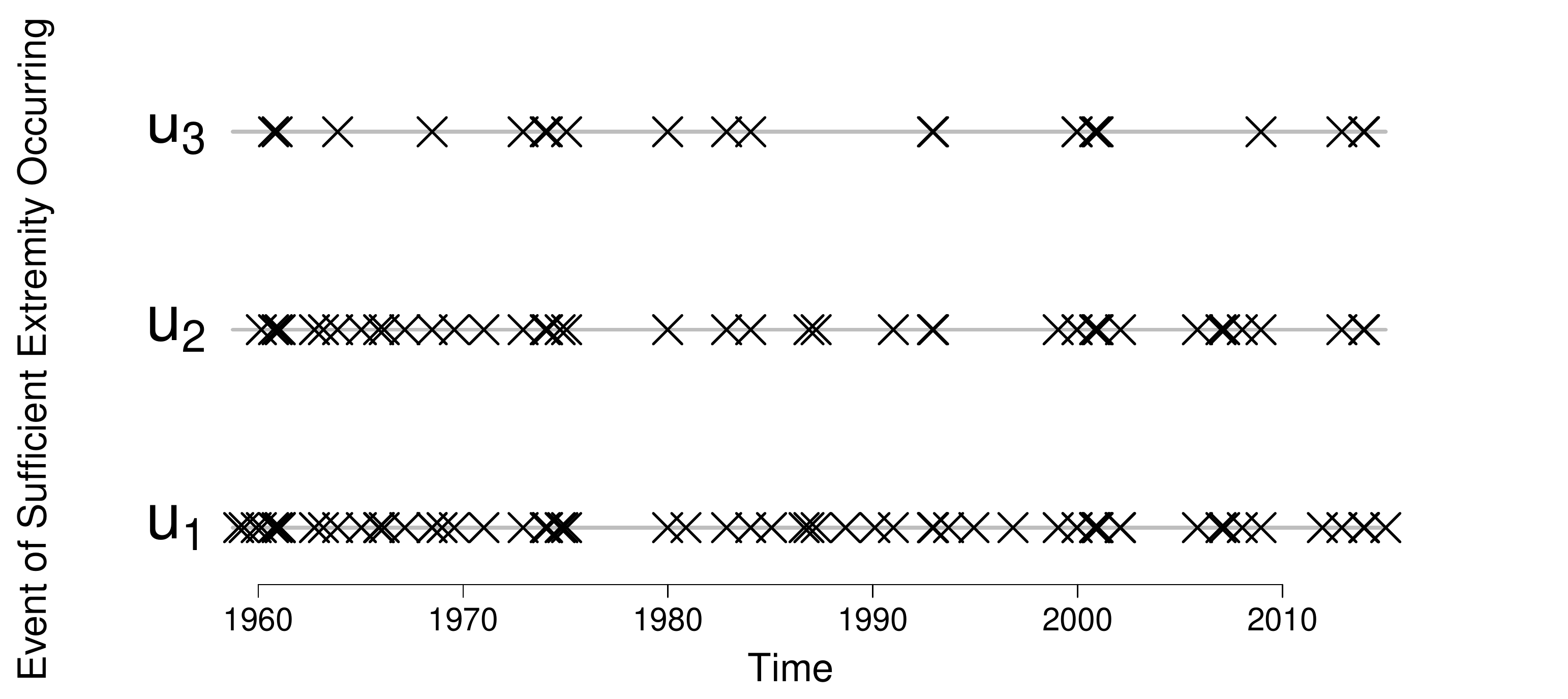}
}
\subfloat[][]{\label{fig:WaTi}
\includegraphics[width=7cm,height=6cm]{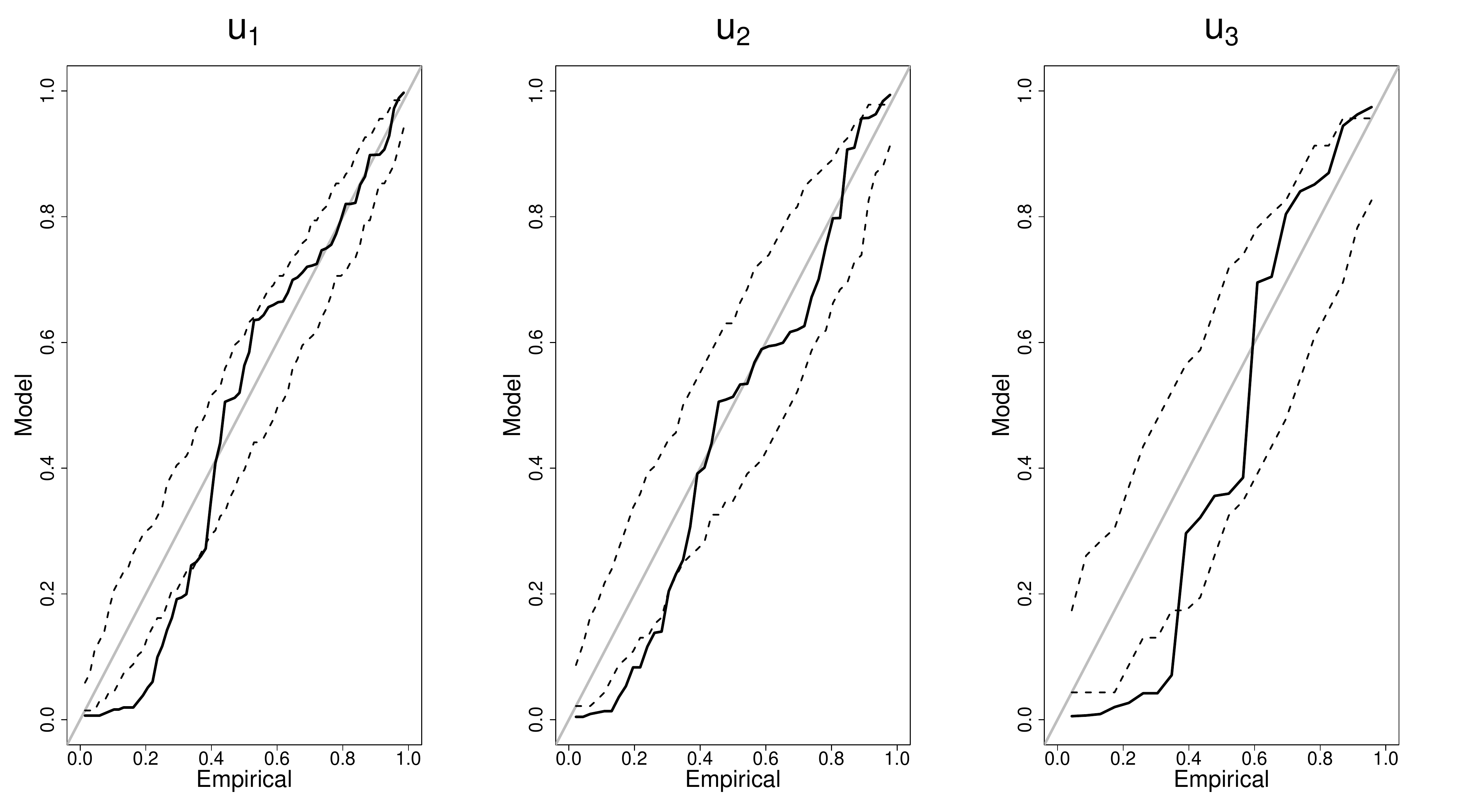}
}
\caption{Exceedance and inter-arrival times for independent storm season events extracted from a south Devon river flow time series above thresholds $u_{1},~u_{2}$ and $u_{3}$ corresponding to 0.7, 0.8 and 0.9 quantiles. (a): Clustering of extreme events with the crosses representing the set of event times. (b): Probability-Probability plots of inter-arrival times for exceedances of $u_{1},~u_{2}$ and $u_{3}$ respectively, 95$\%$ tolerance bands on each plot are derived under the assumption that inter-arrivals are exponentially distributed. The grey line shows the line of equality.}
\label{fig:VisClu}
\end{figure}

The threshold based approaches typically assume that observations are independent and identically distributed (iid). To ensure this is the case, declustering of extreme values is performed to obtain independent extreme events for modelling and inference. The most widely used method to decluster extreme observations is the runs method of \citet{Ferro2003}. The runs method predetermines an event window, $w$, and periods with at least $w$ consecutive observations below the threshold are deemed to define separate independent events. Only the cluster maximum is extracted from each event, to produce a set of independent extreme observations. However, for a declustered river flow time series for a south Devon catchment focussing only on the storm season data, Figure~\ref{fig:ClExt} shows that, over a range of thresholds, clusters of extreme events still exist. If the events were iid, the occurrence times would be well approximated by a Poisson process with exponentially distributed inter-arrival times. The assumption of an exponential distribution is assessed in Figure \ref{fig:WaTi} by use of P-P plots. Figure \ref{fig:WaTi} shows that the inter-arrival times do not follow an exponential distribution, instead the data have a greater probability of short waiting times than is expected under an exponential distribution, and this finding arises for all of the three thresholds choices, that are illustrated in Figure \ref{fig:VisClu}. 

We will show that this clustering of independent events can be described by local non-stationarity, a local change in the marginal distribution of the process. Ignoring this feature leads to biased return period estimates and an over-optimistic assessment of risk following an extreme event. In many cases, local non-stationarity might be linked to changes in climatic covariate values. Previous studies have shown that metrics such as the North Atlantic Oscillation (NAO) or the Southern Oscillation Index (SOI) influence the weather conditions in their respective spatial regions \citep{Trigo2002, Hurrell2003}. Recent research has also focused on the influence of atmospheric rivers (defined as a concentrated narrow channel of heavy vapour) and their influence on winter flooding in the United Kingdom \citep{Lavers2013}, which has the potential to identify different covariate effects by spatial location. 

Detecting how a covariate affects the response (such as river level) can help to improve estimates of return levels and reduce the uncertainty in parameter estimates \citep{Coles2001}. In some cases, there is knowledge of the type of covariates that affect the process of the interest but the data on the covariate are unavailable, or we may have the right covariates but we are unsure of the appropriate functional form of them. Therefore, we adopt similar methods to \citet{Eastoe2010} and \citet{Eastoe2018} to account for the unavailable covariates, through the inclusion of random effects in the parameters of the extreme value models. These approaches are particularly beneficial for accounting for presumed large-scale climatic covariates at small environmental scales, where detecting meaningful relationships is particularly difficult. Accounting for unavailable covariate information through random effect modelling can also improve marginal estimates and reduce the uncertainty in estimates of return values through using more physically realistic statistical models to capture the behaviour of river flow data. 


This paper will present a novel measure to help inform about short-term risk assessment from local non-stationarity; it will provide additional information to the return period which integrates out the local non-stationarity. The proposed measure characterises the heightened short-term risk of extreme events given the occurrence of a $T$ year event at time $t$ in a season. It is similar in interpretation to the definition of relative risk from epidemiology. The risk measure is $R^{(t)}_{T}$ with $0< R^{(t)}_{T}< \infty$: if $R^{(t)}_{T}=1$ there is no change in risk, if $R^{(t)}_{T}>1~(R^{(t)}_{T}<1)$ there is an increased (decreased) risk of observing another extreme event for the remainder of the season. Estimates of this new measure can be derived from our models of local non-stationarity of extreme events. The risk measure is illustrated with both simulated and observed data.




The structure of the paper is as follows: Section \ref{sec:EVT} defines the required underlying univariate extreme value theory used to define the risk measure. Section \ref{sec:RMea} presents the risk measure methodology and this is illustrated through simulation studies in Section \ref{sec:NumIll}. The short-term risk measure is then illustrated in Section \ref{sec:CaSt} in the context of a case study using peak river flow data from Devon. We also provide a practical interpretation of how this risk measure can be used for short-term risk assessment to convey more clearly the reoccurrence chance of extreme events and to help clarify the misinterpretation of an event's return period. 
 
\section{Univariate extreme value theory}\label{sec:EVT}
\subsection{Non-homogeneous Poisson process}\label{sec:AMargModel}
Consider an iid sequence of random variables $Z_{1},\dots,Z_{n}$, and let $M_{n}=\max\{Z_{1},\dots,Z_{n}\}$. If the distribution of the linearly normalised variable $(M_{n}-b_{n})/a_{n}$, $a_{n}>0$, converges to a non-degenerate distribution as $n~\rightarrow~\infty$, this non-degenerate distribution $G$ must be the generalised extreme value distribution (GEV) \citep{Coles2001}, i.e.,
\vspace{-0.7cm}
\begin{center}
\begin{equation}
\mathbb{P}\left(\frac{M_{n}~-~b_{n}}{a_{n}}~\leq~z\right)~\rightarrow~G(z),
\label{eq:Prch1}
\end{equation}
\end{center}
where $G(z)$ takes the following form,
\vspace{-0.7cm}
\begin{center}
\begin{equation}
G(z)=\exp\left\{-\left[1+\xi\left(\frac{z-\mu}{\sigma}\right)\right]^{-\frac{1}{\xi}}_{+} \right\},  
\label{eq:GzCH2}
\end{equation}
\end{center}
with the notation $\left[z\right]_{+}=\max\left\{z,0\right\}$. The GEV distribution function \eqref{eq:GzCH2} is defined through three parameters; location $\mu\in\mathbb{R}$, scale $\sigma>0$ and shape $\xi\in\mathbb{R}$. We may be interested in the maximum up to a scaled time $t\in(0,1]$, i.e., $M_{1:\left\lfloor tn\right\rfloor}$, where here and subsequently $M_{i:j}=\max\{Z_{i},\ldots,Z_{j}\}$ for $i<j$, and $\lfloor x \rfloor$ denotes the integer part of $x$, for example the index $\left\lfloor tn\right\rfloor$ is the number of these $n$ random variables that have been observed up until scaled time $t$. The distribution of the maximum $M_{1:\left\lfloor tn\right\rfloor}$ also converges to a non-degenerate distribution as $n~\rightarrow~\infty$
\vspace{-0.7cm}
\begin{center}
\begin{equation}
\mathbb{P}\left(\frac{M_{1:\left\lfloor tn\right\rfloor}~-~b_{n}}{a_{n}}~\leq~z\right)~\rightarrow~G^{t}(z),
\label{eq:Prtch1}
\end{equation}
\end{center}
with $G$ given by expression \eqref{eq:GzCH2}. The point process representation is an extension of limit \eqref{eq:Prch1} from maxima to all large values and requires the same asymptotic assumptions. Consider a sequence of point processes $P_{n}$, defined on $\mathbb{R}^2$, with 
\vspace{-0.8cm}
\begin{center}
\begin{equation}
P_{n}~=~\left\{\left(\frac{i}{(n+1)},\frac{Z_{i}~-~b_{n}}{a_{n}}\right):~i~=~1,\dots,n\right\},\nonumber
\end{equation}
\end{center}
that converge to a non-homogeneous Poisson process $P$. The scaling of $P_{n}$ is chosen so that the first component, time, is scaled to the interval $(0,1)$ (essentially a normalised temporal index for the random variable $Z$). The second component of $P_{n}$, the maximum of the normalised points, converges in distribution to a non-degenerate limit, given by equation \eqref{eq:Prch1}.
The point process $P_{n}\rightarrow P$ as $n\rightarrow\infty$, on the set $[0,1]\times(u,\infty)$ for all $u<~z$ with $u$ defined such that $u=\sup\left\{z: G(z)~=0\right\}$, for $G(z)$ in equation \eqref{eq:GzCH2}, where $P$ is a non-homogeneous Poisson process with intensity of the form
\vspace{-0.8cm}
\begin{center}
\begin{equation}
\lambda(z,t)~=\frac{1}{\sigma}\left[1+\xi\left(\frac{z-\mu}{\sigma}\right)\right]^{-\frac{1}{\xi}-1}_{+}.\nonumber
\label{eq:lint}
\end{equation}
\end{center}
The integrated intensity for the set $A_{u}~=~[t_{1},t_{2}]\times[u,\infty]$ is
\vspace{-0.8cm}
\begin{center}
\begin{equation}
\Lambda(A_{u})~=~(t_{2}-t_{1})\left[1+\xi\left(\frac{u-\mu}{\sigma}\right)\right]^{-\frac{1}{\xi}}_{+},\nonumber
\label{eq:Lambfa}
\end{equation}
\end{center}
i.e., $\Lambda(A_{u})$  is the expected number of points of $P$ in the set $A_{u}$.  

In a modelling context, the asymptotic limit for the point process representation is assumed to hold for large $n$, i.e., above a high threshold $u$ with the normalising constants $a_{n}>0$ and $b_{n}$ absorbed into the location and scale parameters of the non-homogeneous Poisson process. In the stationary case, the likelihood \citep{Coles2001} for the points that are above the threshold $u$, denoted by $\mathbf{z}_{1:n_{u}}=\left(z_{1},\ldots,z_{n_{u}}\right)$, is
\vspace{-0.7cm}
\begin{center}
\begin{equation}
L(\mu,\sigma,\xi;\mathbf{z}_{1:n_{u}})~\propto~\exp\left\{-n_{y}\left[1+\xi\left(\frac{u-\mu}{\sigma}\right)\right]^{-1/\xi}_{+}\right\}~\prod^{n_{u}}_{i=1}~\frac{1}{\sigma}\left[1+\xi\left(\frac{z_{i}-\mu}{\sigma}\right)\right]^{-\frac{1}{\xi}-1}_{+}
\label{eq:PPCH2Like}
\end{equation}
\end{center}
where $n_{y}$ is the number of years of data ($n_{y}$ could also be the number of blocks, for example the number of winters). With this choice of $n_{y}$, the annual (equivalently block) maximum follows a GEV$(\mu,\sigma,\xi)$ distribution. Inference is performed by using maximum likelihood estimation.   

For standard risk assessments we need to estimate the $T$ year return values denoted by $z_{T}$, defined in Section \ref{sec:Intro}. In the stationary case, we estimate $z_{T}$ by
\vspace{-0.8cm}
\begin{center}
\begin{equation}
\hat{z}_{T}=\hat{\mu}+\frac{\hat{\sigma}}{\hat{\xi}}\left\{\left[-\log\left(1-\frac{1}{T}\right)\right]^{-\hat{\xi}}-1\right\},
\label{eq:rv}
\end{equation}
\end{center}
where $(\hat{\mu},\hat{\sigma},\hat{\xi})$ are estimates obtained by maximising likelihood \eqref{eq:PPCH2Like}.
\subsection{Observed covariates}\label{sec:ObsCov}
In many cases, the distribution of extreme values will be dependent on covariates. This results in the points of the non-homogeneous Poisson process now being independent but non-identically distributed. The covariate space instead of the time space is now considered and as a result points are no longer assumed to be observed uniformly across the space. We initially consider the case where the covariate changes at every observation of $Z_{t}$. Covariate effects can be accounted for in all three parameters of the Poisson process. We focus on a scalar covariate $S$ with corresponding density function $h(s)$. If the non-homogeneous Poisson process then has parameters $\left\{\mu(s),\sigma(s),\xi(s)\right\}$ when $S=s$, the corresponding intensity is
\vspace{-0.6cm}
\begin{center}
\begin{equation}
\lambda(z,s)=n_{y}\frac{h(s)}{\sigma(s)}\left[1+\xi(s)\left(\frac{z-\mu(s)}{\sigma(s)}\right)\right]^{-\frac{1}{\xi(s)}-1}_{+},\nonumber
\end{equation}
\end{center}
and the integrated intensity on $B_{u}=[-\infty,\infty]\times(u,\infty)$, is
\vspace{-0.6cm}
\begin{center}
\begin{equation}
\Lambda(B_{u})=n_{y}\int^{\infty}_{-\infty}\left[1+\xi(s)\left(\frac{u-\mu(s)}{\sigma(s)}\right)\right]^{-\frac{1}{\xi(s)}}_{+}h(s)\mbox{d}s.
\label{eq:covII}
\end{equation}
\end{center}
The integrated intensity, stated in equation \eqref{eq:covII}, is evaluated through the use of numerical integration as typically closed form expressions cannot be obtained. In particular a kernel density estimate for $h(s)$ using all marginal $S$ data is required.

In the presence of non-stationarity due to covariate $S$, the likelihood in \eqref{eq:PPCH2Like} becomes
\vspace{-0.7cm}
\begin{center}
\begin{equation}
L\propto\exp\left\{-n_{y}\int^{\infty}_{-\infty}\left[1+\xi(s)\left(\frac{u-\mu(s)}{\sigma(s)}\right)\right]^{-\frac{1}{\xi(s)}}_{+}h(s)\mbox{d}s\right\}\times\prod^{n_{u}}_{i=1}~\frac{1}{\sigma(s_{i})}\left[1+\xi(s_{i})\left(\frac{z_{i}-\mu(s_{i})}{\sigma(s_{i})}\right)\right]^{-\frac{1}{\xi(s_{i})}-1}_{+},\nonumber
\end{equation}
\end{center} 
where $L=L(\boldsymbol{\mu},\boldsymbol{\sigma},\boldsymbol{\xi};\mathbf{z}_{1:n_{u}},\mathbf{s}_{1:n_{u}})$, with $(\boldsymbol{\mu},\boldsymbol{\sigma},\boldsymbol{\xi})$ being the parameters of $\mu(s),\sigma(s),\xi(s)$ and $\mathbf{s}_{1:n_{u}}=\left(s_{1},\ldots,s_{n_{u}}\right)$ are the covariates associated with extreme values $\mathbf{z}_{1:n_{u}}$. Here we can drop the $h(s)$ terms in the product as they do not involve $\boldsymbol{\mu},~\boldsymbol{\sigma}$ and $\boldsymbol{\xi}$. The standard method to incorporate covariates is to use linear models, with appropriate link functions \citep{Davison1990}, however other more flexible methods such as generalised additive models or splines can also be incorporated into the parameters \citep{Chavez2005,Yee2007}. Regardless of the method being used to model $\mu(s),\sigma(s)$ and $\xi(s)$, the $T$ year return level $z_{T}$ solves the equation 
\vspace{-0.5cm}
\begin{center}
\begin{equation}
\exp\left\{-\int^{\infty}_{-\infty}\left[1+\xi(s)\left(\frac{z_{T}-\mu(s)}{\sigma(s)}\right)\right]^{-\frac{1}{\xi(s)}}_{+}h(s)\mbox{d}s\right\}=1-\frac{1}{T}.
\label{eq:CovRP}
\end{equation}
\end{center}
where the effect of the covariate $S$ is integrated out. 

A special case of a covariate relationship that we will consider is when both the covariate and its effect on the non-homogeneous Poisson process remain constant in each year (or equivalently during a season). Thus for the $i$th of the $n_{y}$ years we have covariate $s_{i}$ and the $n_{u}$ exceedances of $u$ being $\mathbf{z}_{1:n_{u}}=\left\{z_{ij},~j=1,\ldots,n_{u}(i)\right\}$, where $n_{u}(i)$ is the number of exceedances in year $i$ with $\sum_{i}n_{u}(i)=n_{u}$. The likelihood $L=L(\boldsymbol{\mu},\boldsymbol{\sigma},\boldsymbol{\xi};\mathbf{z}_{1:n_{u}},\mathbf{s}_{1:n_{y}})$ is then
\vspace{-0.7cm}
\begin{center}
\begin{eqnarray}
L&\propto&\prod^{n_{y}}_{i=1}\left(\exp\left\{-\left[1+\xi(s_{i})\left(\frac{u-\mu(s_{i})}{\sigma(s_{i})}\right)\right]^{-\frac{1}{\xi(s_{i})}}_{+}\right\}\prod^{n_{u}(i)}_{j=1}~\frac{1}{\sigma(s_{i})}\left[1+\xi(s_{i})\left(\frac{z_{ij}-\mu(s_{i})}{\sigma(s_{i})}\right)\right]^{-\frac{1}{\xi(s_{i})}-1}_{+}\right).\nonumber\\
&\propto&~\exp\left\{-\sum^{n_{y}}_{i=1}\left[1+\xi(s_{i})\left(\frac{u-\mu(s_{i})}{\sigma(s_{i})}\right)\right]^{-\frac{1}{\xi(s_{i})}}_{+}\right\}\prod^{n_{y}}_{i=1}\prod^{n_{u}(i)}_{j=1}~\frac{1}{\sigma(s_{i})}\left[1+\xi(s_{i})\left(\frac{z_{ij}-\mu(s_{i})}{\sigma(s_{i})}\right)\right]^{-\frac{1}{\xi(s_{i})}-1}_{+}.
\nonumber
\end{eqnarray}
\end{center}

\subsection{Unavailable covariates}\label{sec:UnObsCov}
It will not always be possible to obtain an appropriate covariate $S$ as required by the models in Section~\ref{sec:ObsCov}. In this situation, we want to account for covariates without explicitly stating their value. This leads to the adoption of unavailable covariates, otherwise known as random effects \citep{Laird1982}, into the model. Random effects have been adopted to capture unexplained behaviour in a number of previous applications of extreme value theory \citep{Cooley2007, Cooley2010}. Similar methods were presented by \citet{Eastoe2018} for the generalised Pareto distribution, however this is the first time that these models have been presented for the non-homogeneous Poisson process. As with covariates, these random effects can be incorporated into all three parameters of the non-homogeneous Poisson process and can be different for all parameters. We assume that these random effects remain constant within each block of time (e.g., years) $i=1,\ldots,n_{y}$ and are iid over blocks. The collection of all random effects is denoted by $\mathbf{r}=\left(\mathbf{r}_{\mu},\mathbf{r}_{\sigma},\mathbf{r}_{\xi}\right)$ with $\mathbf{r}_{\mu}=(r_{\mu,1},\ldots,r_{\mu,n_{y}})$, where $r_{\mu,i}$ is the random effect for $\mu$ in year $i$ and similarly for $\mathbf{r}_\sigma$ and $\mathbf{r}_\xi$, and $r_{i,j}$ is independent of $r_{k,l}$ when $(i,j)\neq(k,l)$ for all $i,k=\mu,\sigma,\xi$ and $j,l=1,\ldots,n_{y}$. This results in the following parametrisation of the non-homogeneous Poisson process (NHPP) parameters for a given block $i$
\vspace{-0.8cm}
\begin{center}
\begin{eqnarray}
\mu(r_{\mu,i})=\mu_{0}+\mu_{1}r_{\mu,i};~\log\left[\sigma(r_{\sigma,i})\right]=\sigma_{0}+\sigma_{1}r_{\sigma,i};~\xi(r_{\xi,i})=\xi_{0}+\xi_{1}r_{\xi,i},
\label{eq:parset}
\end{eqnarray}
\end{center}
where $\left(\boldsymbol{\mu},\boldsymbol{\sigma},\boldsymbol{\xi}\right)=(\mu_{0},\mu_{1};\sigma_{0},\sigma_{1};\xi_{0},\xi_{1})$. As the parametrisation in equation \eqref{eq:parset} is linear in the random effect, we can assume that the random effects have zero mean and unit variance. For parsimony and computational convenience we assume that they jointly follow the multivariate Normal distribution
\vspace{-0.7cm}  
\begin{center}
\begin{eqnarray}
\left[\begin{matrix}r_{\mu_{i}} \\ r_{\sigma_{i}}  \\ r_{\xi_{i}}\end{matrix}\right]~\sim~\mbox{MVN}\left( \left[\begin{matrix} 0 \\ 0 \\ 0 \end{matrix}\right]~,~\Sigma=\left[\begin{matrix} 1 & \rho_{\mu,\sigma} & \rho_{\mu,\xi}\\
 \rho_{\mu,\sigma} & 1 & \rho_{\sigma,\xi}\\
  \rho_{\mu,\xi}& \rho_{\sigma,\xi} & 1\\
 \end{matrix}\right]\right)\nonumber,
\label{eq:ReDist}
\end{eqnarray}
\end{center}
where $\Sigma$ is the covariance matrix and $\boldsymbol{\rho}=(\rho_{\mu,\sigma},\rho_{\mu,\xi},\rho_{\sigma,\xi})$ represents the correlations between the random effects incorporated into the specific parameters. The single site likelihood with the inclusion of random effects that remain constant over a given block is defined as follows
\vspace{-0.7cm} 
\begin{center}
\begin{eqnarray}
L(\boldsymbol{\mu},\boldsymbol{\sigma},\boldsymbol{\xi},\mathbf{r}_{1:n_{y}},\boldsymbol{\rho};\mathbf{z}_{1:n_{u}})~\propto~\exp\left\{-\sum^{n_{y}}_{i=1}\left[1+\xi(r_{\xi,i})\left(\frac{u-\mu(r_{\mu,i})}{\sigma(r_{\sigma,i})}\right)\right]^{-\frac{1}{\xi(r_{\xi,i})}}_{+}\right\}\nonumber\\\times\prod^{n_{y}}_{i=1}\left\{\prod^{n_{u}(i)}_{j=1}~\frac{1}{\sigma(r_{\sigma,i})}\left[1+\xi(r_{\xi,i})\left(\frac{z_{ij}-\mu(r_{\mu,i})}{\sigma(r_{\sigma,i})}\right)\right]^{-\frac{1}{\xi(r_{\xi,i})}-1}_{+}\right\}\phi(r_{\mu,i},r_{\sigma,i},r_{\xi,i};\Sigma),
\label{eq:ILRe}
\end{eqnarray}
\end{center}

where $\phi$ is the probability density function of the standard multivariate Normal with covariance matrix $\Sigma$, so the density of the random effects is now included. 

Unlike in Section \ref{sec:ObsCov}, obtaining maximum likelihood estimates for the parameters and random effects in \eqref{eq:ILRe} is difficult. As a result, the methods of \citet{Eastoe2018} are adopted. We use a Bayesian framework and estimate the parameters using Markov chain Monte Carlo (MCMC) methods with a Metropolis-Hastings random walk \citep{Gilks1995}. The adoption of Bayesian methods allows us to incorporate prior information, which can be ascertained from domain experts as described in \citet{Coles1996}. In order to obtain convergence of the MCMC chains, the distributions from which parameters are proposed are tuned by using an adaptive algorithm until the optimal acceptance rate is reached \citep{Roberts2009}. Standard visual assessments, such as trace plots, are performed to assess the convergence of the chains.

For a historical block $i$ for which we have an estimate of the random effects, we can determine a block $i$ specific return level $z_{T,i}$ through solving equation \eqref{eq:RERP} 
\vspace{-1cm}
\begin{center}
\begin{equation}
\exp\left\{-\left[1+\hat{\xi}(\hat{r}_{\xi,i})\left(\frac{z_{T,i}-\hat{\mu}(\hat{r}_{\mu,i})}{\hat{\sigma}(\hat{r}_{\sigma,i})}\right)\right]^{-\frac{1}{\hat{\xi}(\hat{r}_{\xi,i})}}_{+}\right\}=1-\frac{1}{T},
\label{eq:RERP}
\end{equation}
\end{center}
with the parameters and the random effect evaluated at the posterior estimates. If required, we can still obtain a single cross-year return year level curve by integrating over random effects
\vspace{-0.8cm}
\begin{center}
\begin{equation}
\int_{\mathbb{R}^{3}}\exp\left\{-\left[1+\hat{\xi}(r_{\xi})\left(\frac{z_{T}-\hat{\mu}(r_{\mu})}{\hat{\sigma}(r_{\sigma})}\right)\right]^{-\frac{1}{\hat{\xi}(r_{\xi})}}_{+}\right\}\phi(\mathbf{r};\hat{\Sigma})\mbox{d}\mathbf{r}=1-\frac{1}{T}
\label{eq:AvgRE}
\end{equation}
\end{center}
and solving to obtain an estimate of $z_{T}$.
\section{Short-term risk measure}\label{sec:RMea}
\subsection{Risk measure for threshold exceedance data}\label{sec:ViRe}
We propose a short-term risk measure that is updated for the rest of a block using information on the size of the largest event that has been observed up to this point in the block. Consider a window of normalised time $[0,1]$, for which the covariate remains constant and consider exceedances of a threshold \textit{u}; above which the statistical model holds.

\begin{figure}[!h]
\centering
\vspace{-0.2cm}
\subfloat[]{\label{fig:NHPP}
\includegraphics[width=8cm,height=6cm]{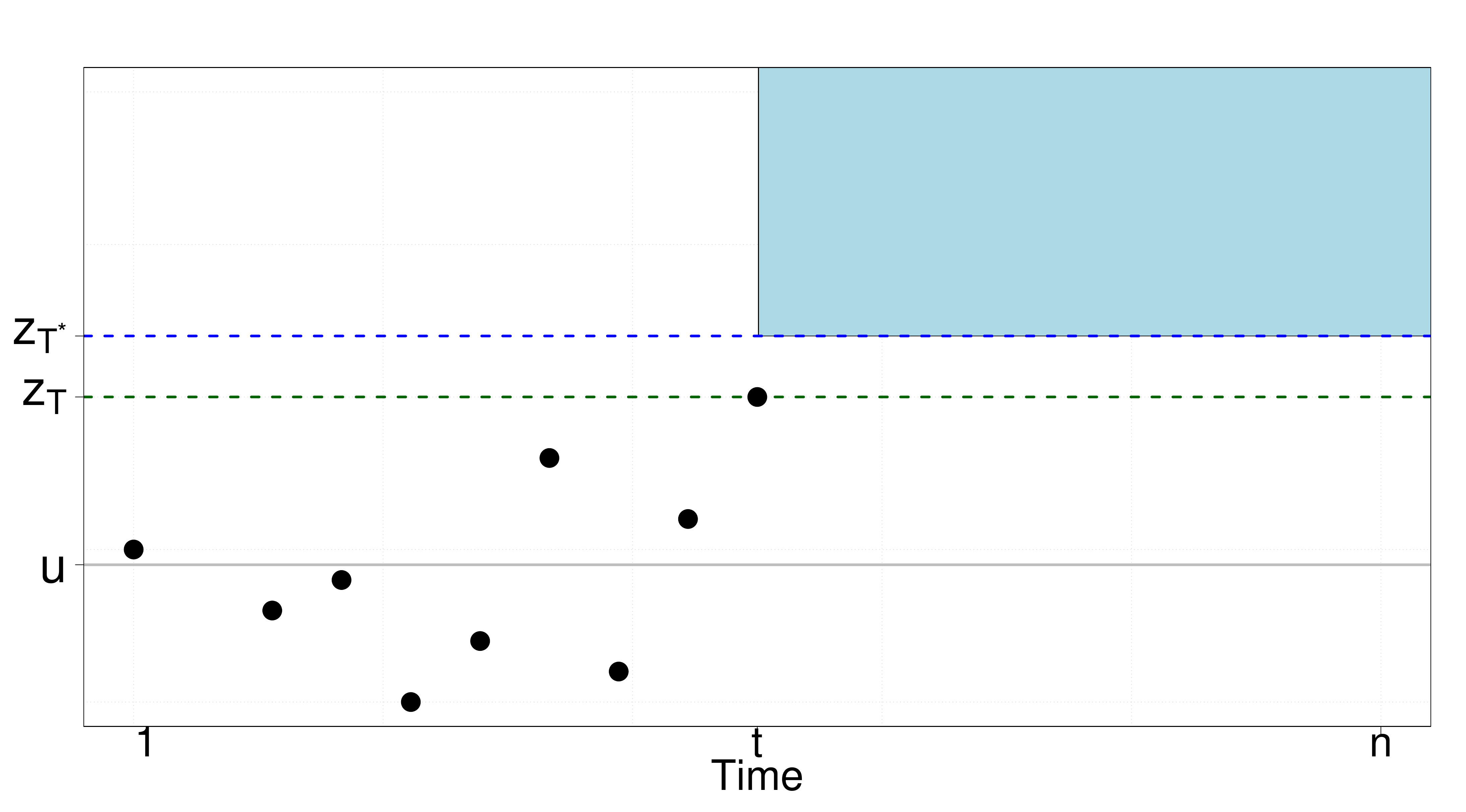}
}
\subfloat[]{\label{fig:AltRM}
\includegraphics[width=8cm,height=6cm]{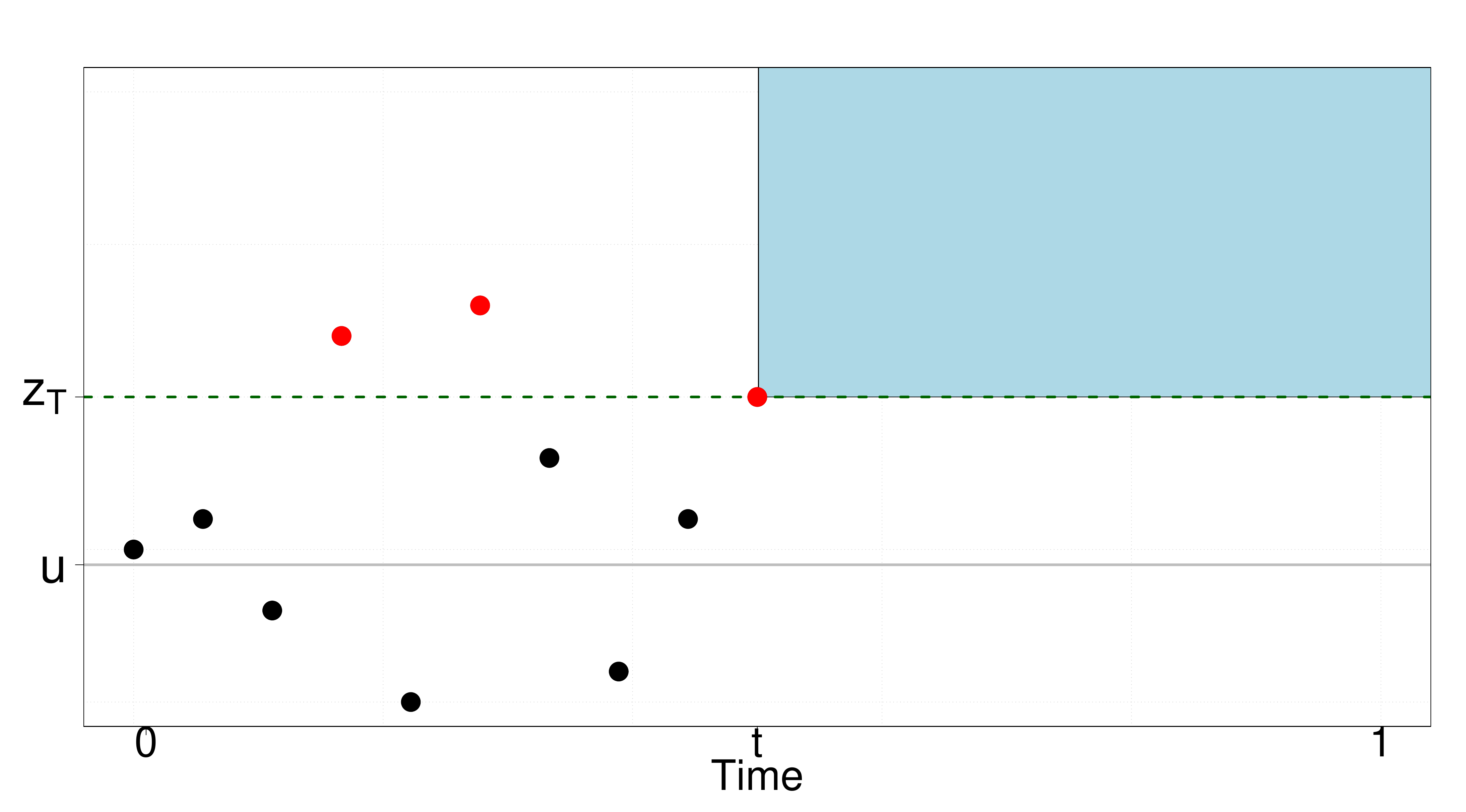}
}
\caption{(a): Visualisation of the risk measure for the threshold exceedance modelling approach with $Z_{t}=z_{T}$, where $z_{T}$ has a $T$ year return period. The blue shaded area represents the region $B_{t,z_{T^{*}}}=[t,1]\times[z_{T^{*}},\infty)$. (b): Visualisation of an alternative risk measure (discussed in Section \ref{sec:Dis}), whereby we are interested in the probability of any points being observed in the region $B_{t,z_{T}}=[t,1]\times[z_{T},\infty)$ given that a number of points were above the level $z_{T}$ before time $t$.}
\label{fig:RM}
\end{figure}

For a given time series suppose that a $T$-year event of size $z_{T}$ is observed at time $t$. As a result of this event $z_{T}$ we want to determine for the remainder of time period $(t,1]$, the probability of observing an event of size greater than $z_{T^{*}}$, with $z_{T^{*}}>u$, corresponding to a $T^{*}$ year return period event. We denote this by $R^{(t)}_{T,T^{*}}$. Most often we envisage that interest will be when $T^{*}=T$, i.e., any future events are as rare as the previously largest event, and we denote this by $R^{(t)}_{T}$. A visualisation of $R^{(t)}_{T,T^{*}}$ is given in Figure \ref{fig:NHPP} with $Z_{t}=z_{T}$, and with $z_{T^{*}}>z_{T}$ so the return period $T^{*}>T$. 

Given that $M_{1:\lfloor tn \rfloor}=z_{T}>u$ we are interested in determining for the remaining observations $Z_{\lfloor tn \rfloor+1},\ldots,Z_{n}$ the probability that $M_{\lfloor tn \rfloor+1:n}=\max\left\{Z_{\lfloor tn \rfloor+1},\ldots,Z_{n}\right\}>z_{T^{*}}$ asymptotically as $n\rightarrow\infty$. In terms of the Poisson process this is equal to saying that we are interested in the probability of there being at least one point in the set $B_{t,z_{T^{*}}}=[t,1]\times[z_{T^{*}},\infty)$ given $M_{1:\lfloor tn \rfloor}=z_{T}$, see Figure \ref{fig:NHPP}. This probability is compared to the marginal probability of $M_{\lfloor tn \rfloor +1:n}>z_{T^{*}}$ without conditioning on the maximum value $M_{1:\lfloor tn\rfloor}$ being observed up to time $t$; this allows us to assess the effect of observing an event of size $z_{T}$. We define the measure of short-term risk by 
\vspace{-0.5cm}
\begin{center}
\begin{equation}
R^{(t)}_{T,T^{*}}= \frac{\mathbb{P}(M_{\lfloor tn \rfloor+1:n}>z_{T^{*}}|M_{1:\lfloor tn \rfloor}=z_{T})}{\mathbb{P}(M_{\lfloor tn \rfloor+1:n} > z_{T^{*}})}, 
\label{eq:newRR}
\end{equation}
\end{center}
where $R^{(t)}_{T,T^{*}}\in[0,\infty)$ and if $R^{(t)}_{T,T^{*}}=1$ there is no change in risk, however if $R^{(t)}_{T,T^{*}}>1~(<1)$, there is an increase (decrease) in risk of an event of size $z_{T^{*}}$ occurring. 

\subsection{Derivation of the risk measure under a covariate effect}\label{sec:DerObs}
Knowing that $M_{1:\lfloor tn \rfloor}=z_{T}$ provides information about the value of the covariate $S$ and consequently about how the risk has changed for that specific block of time. The denominator in definition \eqref{eq:newRR} can be calculated by using the GEV parametrisation of the non-homogeneous Poisson process with a covariate $S$ common over the block as follows
\vspace{-0.7cm}
\begin{center}
\begin{eqnarray}
\mathbb{P}(M_{\lfloor tn\rfloor+1:n}\leq z)&=&\int^{\infty}_{-\infty}\mathbb{P}(M_{\lfloor tn \rfloor +1:n}\leq z|S=v)h(v)\mbox{d}v\nonumber\\
			 					 &=&\int^{\infty}_{-\infty}\exp\left\{-(1-t)\left[1+\xi(v)\left(\frac{z-\mu(v)}{\sigma(v)}\right)\right]^{-\frac{1}{\xi(v)}}_{+}\right\}h(v)\mbox{d}v,\nonumber
\label{eq:denRR}
\end{eqnarray}
\end{center}
where $h$ is the density of the covariate and $v$ is a dummy variable for the covariate $S$ which is being integrated over. The conditional probability of $M_{\lfloor tn \rfloor +1:n}$ being above a level $z$ given that the maximum of $M_{1:\lfloor tn \rfloor}$ is $z$ corresponds to the numerator of expression \eqref{eq:newRR} and is calculated as follows,
\vspace{-0.8cm}
\begin{center}
\begin{equation}
\mathbb{P}(M_{\lfloor tn \rfloor +1:n}>z|M_{1:\lfloor tn \rfloor}=z)=\int^{\infty}_{-\infty}\mathbb{P}(M_{\lfloor tn \rfloor +1:n}>z|S=v)h_{S|1:\lfloor tn \rfloor}(v|z)\mbox{d}v,\nonumber
\label{eq:n1}
\end{equation}
\end{center}
where
\vspace{-1cm}
\begin{center}
\begin{equation}
h_{S|1:\lfloor tn \rfloor}(v|z)=\frac{g_{1:\lfloor tn \rfloor|S}(z|v)h(v)}{g_{1:\lfloor tn \rfloor}(z)},
\label{eq:cond}
\end{equation}
\end{center}
where $g_{1:\lfloor tn \rfloor}(z)$ and $g_{1:\lfloor tn \rfloor|S}(z|v)$ are the marginal and conditional on $S$ GEV densities of $M_{1:\lfloor tn \rfloor}$. In order to calculate these densities, we can use the marginal probability that
\vspace{-0.8cm}
\begin{center}
\begin{equation}
\mathbb{P}(M_{1:\lfloor tn \rfloor}<z)=\int^{\infty}_{-\infty}\exp\left\{-t\left[1+\xi(v)\left(\frac{z-\mu(v)}{\sigma(v)}\right)\right]^{-\frac{1}{\xi(v)}}_{+}\right\}h(v)\mbox{d}v\nonumber
\label{eq:Pmax}
\end{equation}
\end{center}
and by differentiation it follows that,
\vspace{-0.8cm}
\begin{center}
\begin{eqnarray}
g_{1:\lfloor tn \rfloor}(z)=\int^{\infty}_{-\infty}g_{1:\lfloor tn \rfloor|S}(z|v)h(v)\mbox{d}v,\nonumber
\label{eq:Ma1xt}
\end{eqnarray}
\end{center}
where
\vspace{-0.8cm}
\begin{center}
\begin{eqnarray}
g_{1:\lfloor tn \rfloor|S}(z|v)=\frac{t}{\sigma(v)}\left[1+\xi(v)\left(\frac{z-\mu(v)}{\sigma(v)}\right)\right]^{-\frac{1}{\xi(v)}-1}_{+}\exp\left\{-t\left[1+\xi(v)\left(\frac{z-\mu(v)}{\sigma(v)}\right)\right]^{-\frac{1}{\xi(v)}}_{+}\right\}.\nonumber
\label{eq:Maxt}
\end{eqnarray}
\end{center}
\subsection{Derivation of the risk measure for an unavailable covariate}\label{sec:DerUnobs}
The risk measure of equation \eqref{eq:newRR} can also be estimated if random effects are incorporated into the parameters. The marginal probability given in equation \eqref{eq:newRR} therefore becomes 
\vspace{-0.8cm}
\begin{center}
\begin{eqnarray}
\mathbb{P}(M_{\lfloor tn \rfloor +1:n}\leq z_{T^{*}})&=&\int_{\mathbb{R}^{3}}\mathbb{P}(M_{\lfloor tn \rfloor+1:n}\leq z_{T^{*}}|\mathbf{R}=\mathbf{r})\phi(\mathbf{r};\Sigma)\mbox{d}\mathbf{r}\nonumber\\
			 					 &=&\int_{\mathbb{R}^{3}}\exp\left\{-(1-t)\left[1+\xi(r_{\xi})\left(\frac{z_{T^{*}}-\mu(r_{\mu})}{\sigma(r_{\sigma})}\right)\right]^{-\frac{1}{\xi(r_{\xi})}}_{+}\right\}\phi(\mathbf{r};\Sigma)\mbox{d}\mathbf{r},\nonumber
\label{eq:reRR}
\end{eqnarray}
\end{center}
where $\mathbf{r}=(r_{\mu},r_{\sigma},r_{\xi})$ and $\phi$ is the multivariate Normal density defined in equation \eqref{eq:ReDist}. The numerator of expression \eqref{eq:newRR} is calculated as follows,
\vspace{-0.8cm}
\begin{center}
\begin{equation}
\mathbb{P}(M_{\lfloor tn \rfloor+1:n}>z_{T^{*}}|M_{1:\lfloor tn \rfloor}=z)=\int_{\mathbb{R}^{3}}\mathbb{P}(M_{\lfloor tn \rfloor+1:n}>z_{T^{*}}|\mathbf{R}=\mathbf{r})\phi_{\mathbf{R}|1:\lfloor tn \rfloor}(\mathbf{r};\Sigma)\mbox{d}\mathbf{r},\nonumber
\label{eq:ren1}
\end{equation}
\end{center}
where
\vspace{-1.2cm}
\begin{center}
\begin{equation}
\phi_{\mathbf{R}|1:\lfloor tn \rfloor}(\mathbf{r};\Sigma|z)=\frac{g_{1:\lfloor tn \rfloor|\mathbf{R}}(z|\mathbf{r})\phi(\mathbf{r};\Sigma)}{g_{1:\lfloor tn \rfloor}(z)},\nonumber
\label{eq:recond}
\end{equation}
\end{center}
and $g_{1:\lfloor tn \rfloor}(z)$ and $g_{1:\lfloor tn \rfloor|\mathbf{R}}(z|\mathbf{r})$ are again the marginal and conditional GEV densities of $M_{1:\lfloor tn \rfloor}$. Finally, through using the same strategy as in Section \ref{sec:DerObs} we have
\vspace{-1cm}
\begin{center}
\begin{eqnarray}
g_{1:\lfloor tn \rfloor}(z)&=&\int_{\mathbb{R}^{3}}g_{1:\lfloor tn \rfloor|\mathbf{R}}(z|\mathbf{r})\phi(\mathbf{r};\Sigma)\mbox{d}\mathbf{r},\nonumber
\label{eq:Maxt}
\end{eqnarray}
\end{center}
where
\vspace{-0.8cm}
\begin{center}
\begin{eqnarray}
g_{1:\lfloor tn \rfloor|\mathbf{R}}(z|\mathbf{r})=\frac{t}{\sigma(r_{\sigma})}\left[1+\xi(r_{\xi})\left(\frac{z-\mu(r_{\mu})}{\sigma(r_{\sigma})}\right)\right]^{-\frac{1}{\xi(r_{\xi})}-1}_{+}\exp\left\{-t\left[1+\xi(r_{\xi})\left(\frac{z-\mu(r_{\mu})}{\sigma(r_{\sigma})}\right)\right]^{-\frac{1}{\xi(r_{\xi})}}_{+}\right\}.\nonumber
\label{eq:Mtaxt}
\end{eqnarray}
\end{center}

\section{Numerical illustration of the risk measure}\label{sec:NumIll}
\begin{figure}[!htbp]
\centering
\includegraphics[width=6cm,height=6cm]{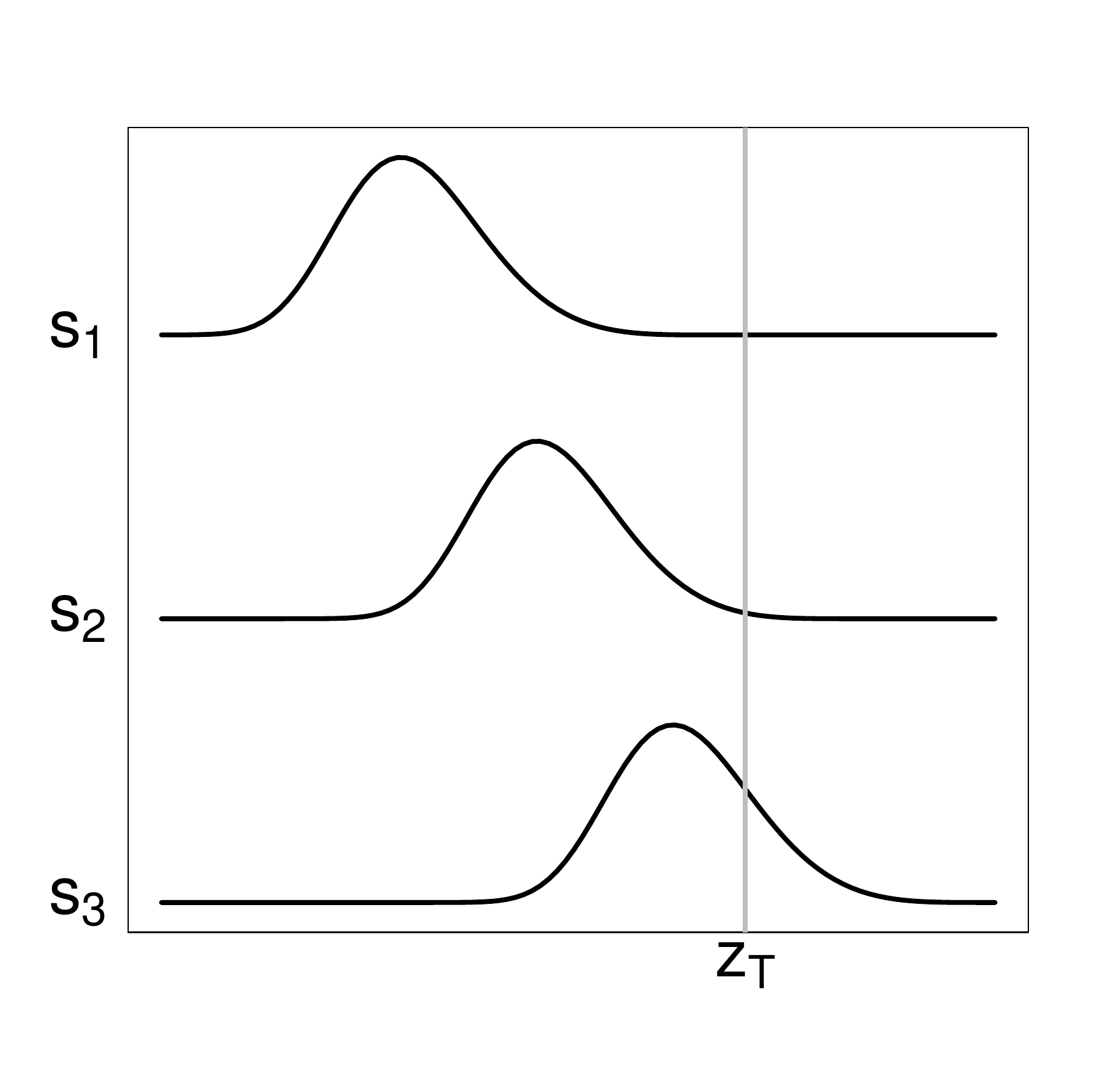}
\caption{Conditional density for a range of $z$, showing the impact of the change in value of the covariate $s$. Three choices of the covariates are shown $s_{1},~s_{2}$ and $s_{3}$, which correspond to $2.5,~50$ and $97.5\%$ quantile of the covariate density. The vertical line corresponds to the value of $z_{T}$.}
\label{fig:REzT}
\end{figure}

To illustrate the ideas behind our proposal for the risk measure, Figure \ref{fig:REzT} shows a simplistic version of the problem when the covariate $S$ can take only three different values $s_{1}$, $s_{2}$ and $s_{3}$ and we observe a large value, the $T$ year return level $z_T$. Here $s_{1}$, $s_{2}$ and $s_{3}$ corresponding to the $2.5,~50$ and $97.5\%$ quantiles of a $\mbox{N}(0,1)$ variable and the distribution of the random variable of interest, $Z$, given $S$ has a negative shape parameter. Hence, $z_T$ is impossible for $Z$ when $S=s_1$. By comparing the conditional densities given that $z_{T}$ has been observed, it is clear that $S=s_{3}$ is the most likely value of $S$. As this value of $S$ is larger than the average $S$ it follows that $R^{(0.4)}_{T,T^{*}}>1$. More generally $S$ has a continuum of possible values, and an observed maximum value of $z_{T}$ in a year so far, will favour some range of values of $S$ over others and hence gives an updated short-term risk for the rest of the year.

The risk measure \eqref{eq:newRR} is illustrated through the case of a linear trend in the location parameter of the $\mbox{NHPP}(\mu(s)=\alpha+\beta s,\sigma(s)=\sigma _{0},\xi(s)=\xi_{0})$ with 30 years' worth of simulated data. The following parametrisation is used, with $\alpha=0,~\beta=2.5,~\sigma_{0}=1.5$ and the positive and negative shape parameter cases $\xi_{0}=\pm 0.2$, with $S\sim\mbox{N}(0,1)$. We assume that the covariate effect only exists in the location parameter and not the scale and shape parameters. The choice of values of the shape parameter is consistent with observed values in the environment.  We are interested in the occurrence of a value above $z_{T^{*}}$ after the 1 in \textit{T} year event and we consider this for $T^{*} \in [1,100]$. We condition on $z_{T}$ being equal to a 1 in 100 year event and the event occurring at $t=0.4$, i.e., $40\%$ of the way into the storm season. 

\begin{figure}[!htbp]
\centering
\vspace{-0.2cm}
\subfloat[]{\label{fig:Pos}
\includegraphics[width=6cm,height=6cm]{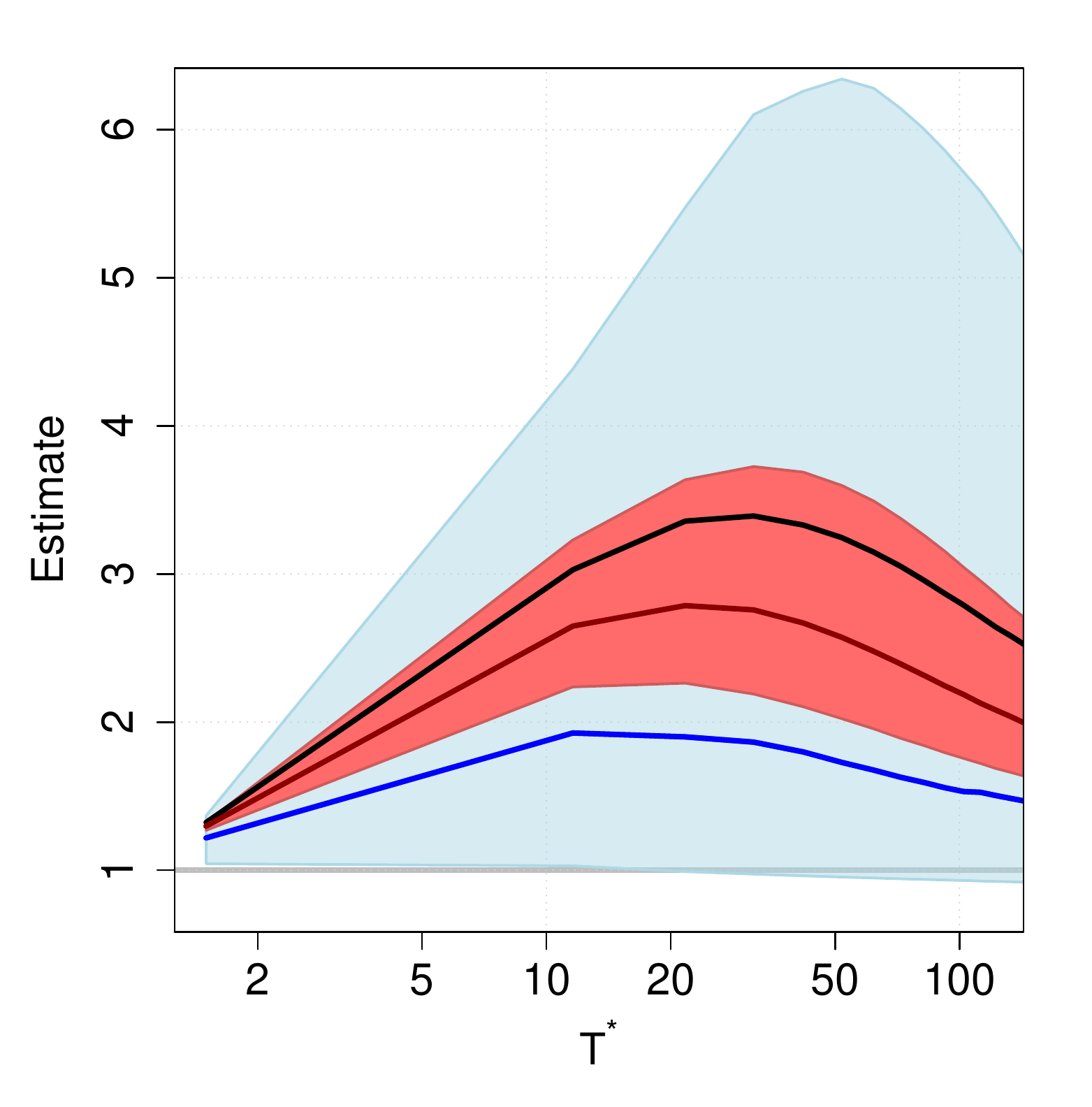}
}
\subfloat[]{\label{fig:Neg}
\includegraphics[width=6cm,height=6cm]{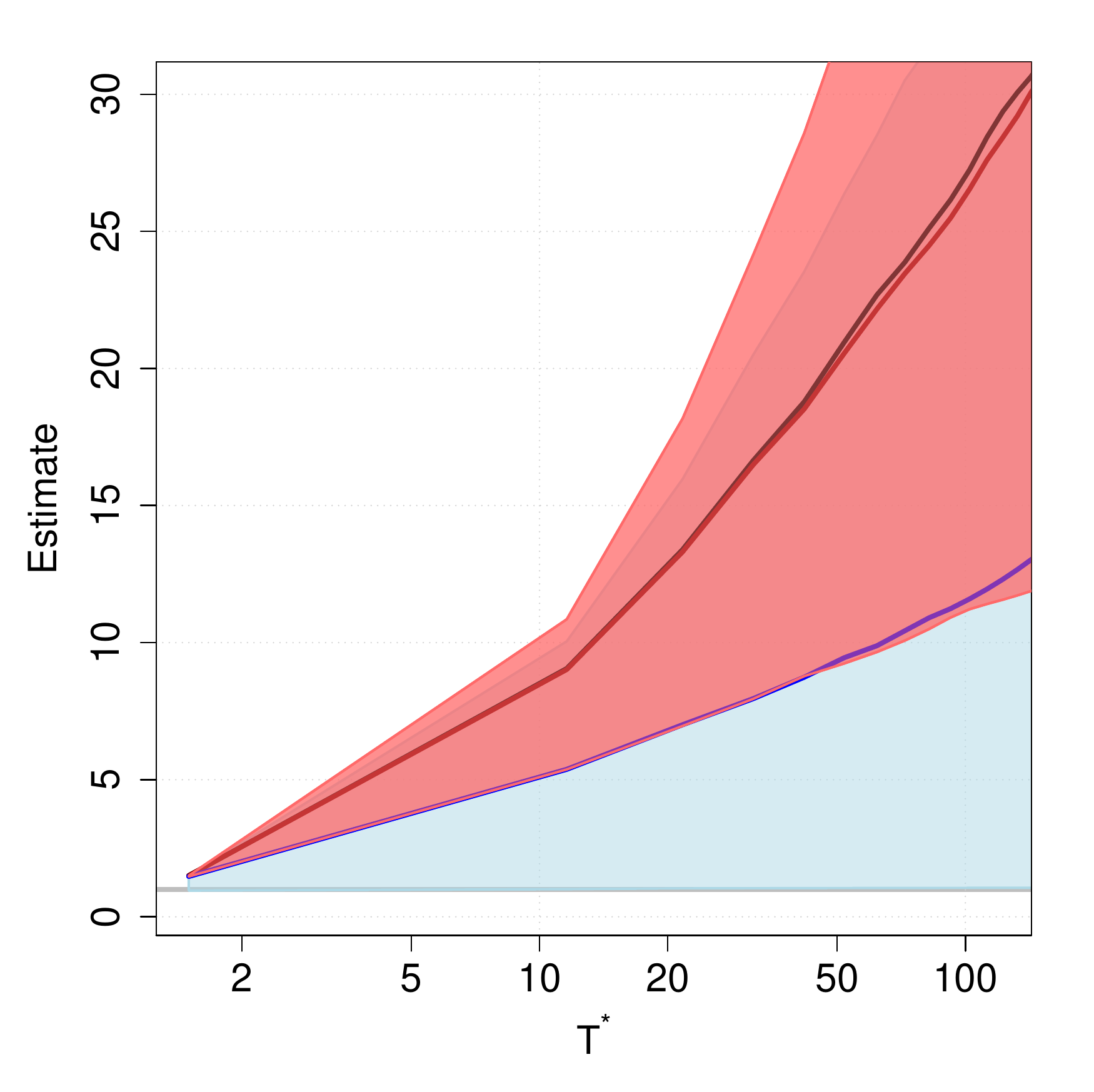}
}
\caption{Estimates of the short-term risk measure $R^{(0.4)}_{100,T^{*}}$ when (a) $\xi>0$ and (b) $\xi<0$. The true values are shown by the black lines: The point estimate of the risk when the covariate is observed (unavailable) are shown by the red (blue) lines with associated 95$\%$ pointwise uncertain intervals given by the shaded sets with identical colour (intervals are overlaid). Uncertainty intervals are calculated from 50 simulated samples of 30 years' worth of data.}
\label{fig:reRR}
\end{figure}

Figure~\ref{fig:reRR} shows that the true short-term risk measure $R^{(0.4)}_{100,T^{*}}$ is above one for both values of $\xi$. When $\xi>0$, the distribution is unbounded so there is always a non-zero probability, however small, of being above a level $z_{T^{*}}$ whatever the covariate. However, the most dramatic effect is seen with $\xi<0$, with the short-term risk measure increasing with level in Figure~\ref{fig:reRR}~\subref{fig:Neg} with a much larger magnitude than when $\xi>0$. For example, the chance of observing a 1 in 50 year event is now more like a 1 in 2 year event. 


We can also estimate the risk measure when the covariate $S$ is either available or unavailable and the parameters of the non-homogeneous Poisson process are to be estimated. In the case where the covariate is available the methods in Sections \ref{sec:ObsCov} and \ref{sec:DerObs} are used. When the covariate $S$ is unavailable but there is evidence that the data are not identically distributed, the methods in Sections \ref{sec:UnObsCov} and \ref{sec:DerUnobs} provide us with estimates of random effects, which represent the unavailable covariate. 

These two (with and without covariates) different estimates of short-term risk in Figure \ref{fig:reRR} are slightly lower than the truth but all estimate an increase in risk and capture the pattern of how $R^{(0.4)}_{100,T^{*}}$ varies with $T^{*}$. When the covariates are unavailable the estimates are lower than when they are available and have larger uncertainty intervals; this is expected as there are a larger number of parameters to estimate. The uncertainty intervals show that both estimation methods give intervals which contain the truth. The estimates in Figure \ref{fig:reRR} show the benefits of using random effect models to account for an unavailable covariate as they give similar risk measure estimates to when the correct covariate is observed. Furthermore, if we had not included random effects here the risk measure would have been estimated to be equal to one, considerably underestimating the associated short-term risk. 

\begin{figure}[!htbp]
\centering
\vspace{-0.2cm}
\subfloat[]{\label{fig:Pos}
\includegraphics[width=6cm,height=6cm]{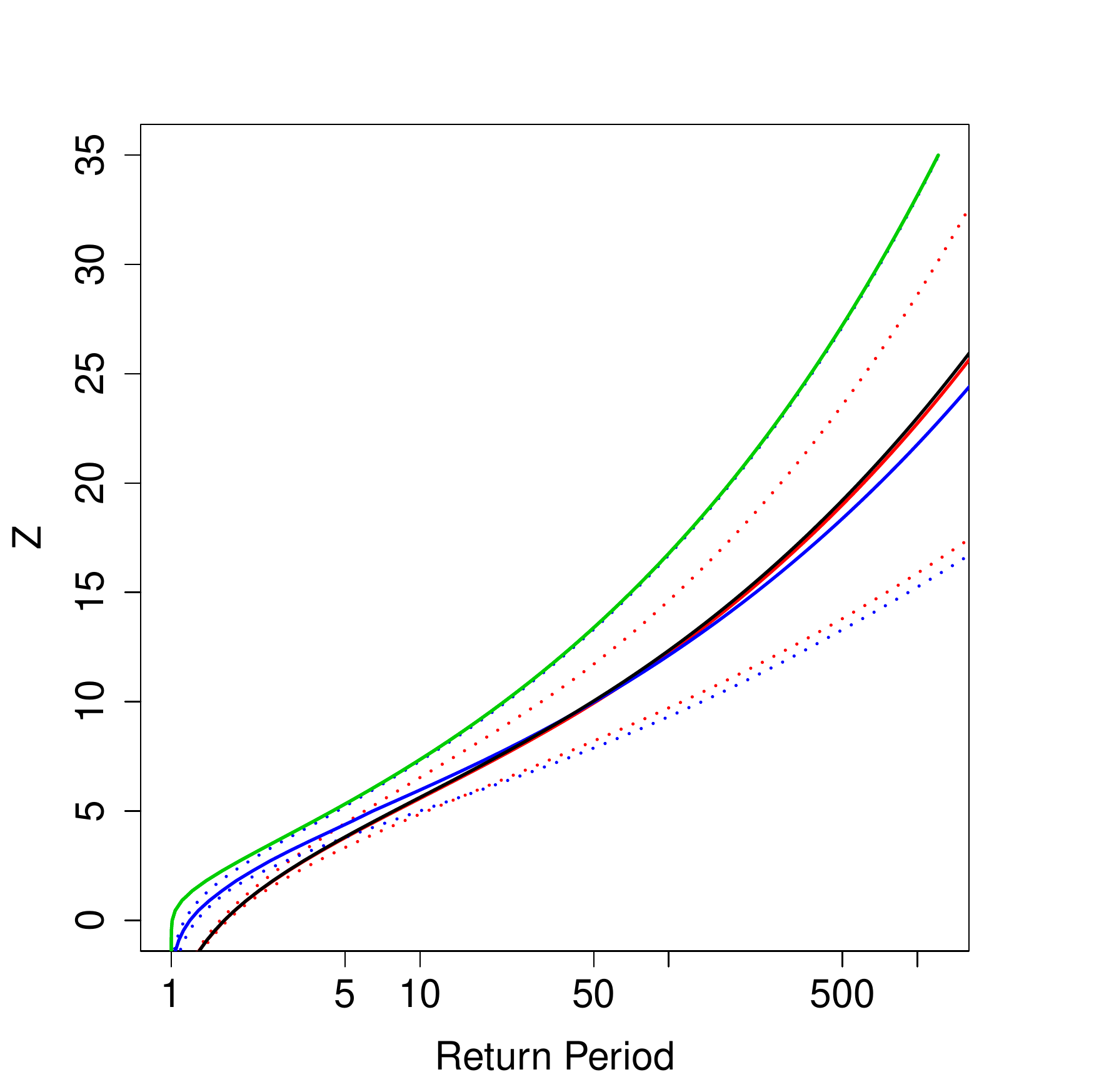}
}
\subfloat[]{\label{fig:Neg}
\includegraphics[width=6cm,height=6cm]{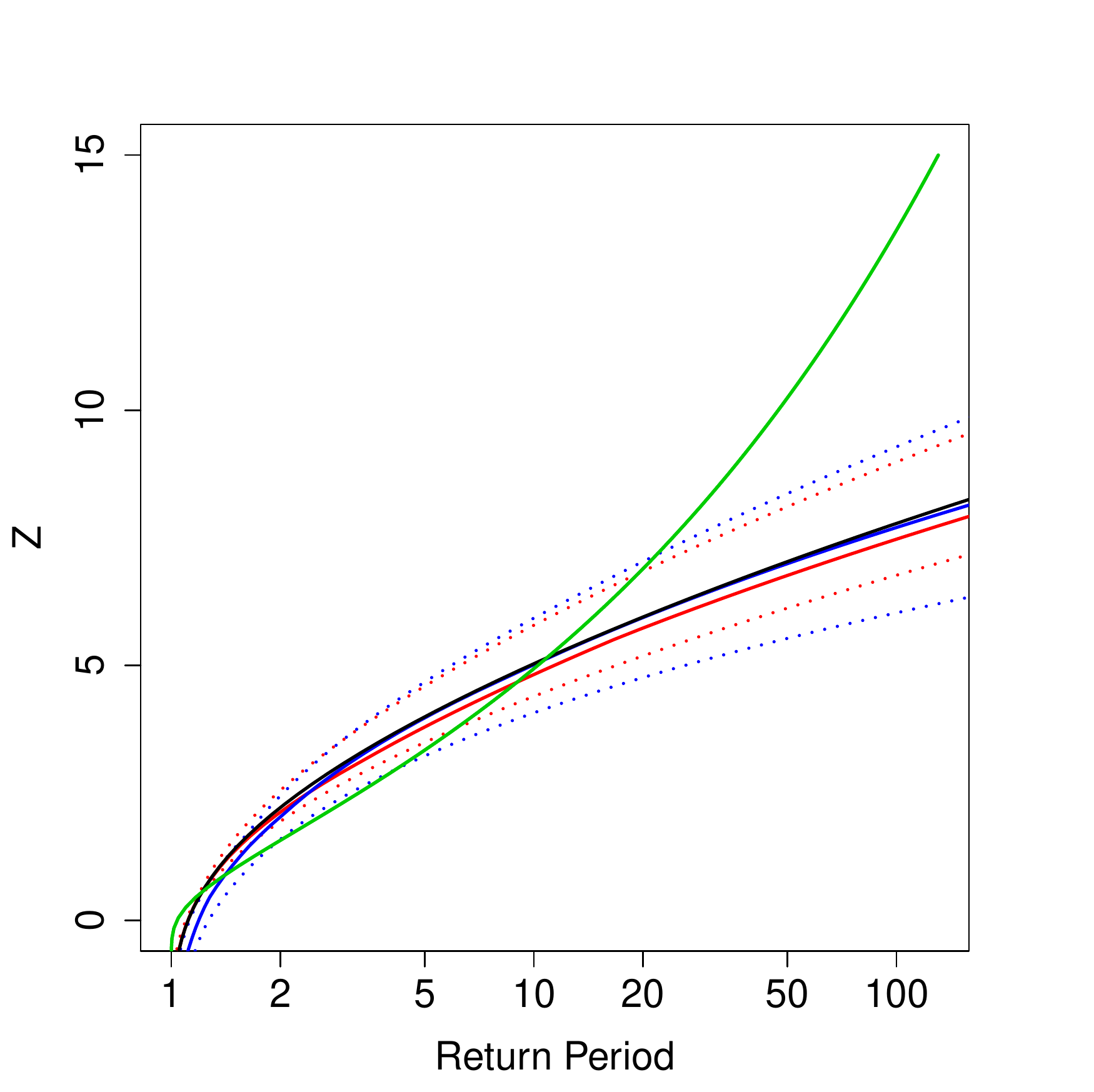}
}
\caption{Comparison of the estimated return levels for different methods (a) $\xi>0$ and (b) $\xi<0$. The true values are shown by the black lines. The cases where the covariate is observed (unavailable) are shown by the red (blue) lines.  Uncertainty intervals are calculated from 10 simulated samples of 30 years' worth of data and are shown as dotted lines in the same colours as the point estimates. The green line shows the estimate if we ignore the existence of the covariate.}
\label{fig:DisMax}
\end{figure}

Comparisons of estimates of the return levels of the annual maxima for a random year within the observation period are given in Figure \ref{fig:DisMax} for the same values of the shape parameter as in Figure \ref{fig:reRR}. The estimates for when the covariates are observed agrees very well with the truth. When the covariates are unavailable the estimates are still good, but naturally they perform slightly less well than if the covariate was observed. In contrast, if the covariate is ignored and an iid GEV distribution is wrongly fitted, this estimate is considerably different, in each case giving far too heavy tails and overestimating return levels because the covariate variation across years is interpreted as variation to be modelled by the GEV. This comparison highlights the importance of accounting for covariates to explain the local non-stationarity of extreme events, as lower and more accurate return levels estimates are achieved. 

The behaviour of the return level estimates are consistent with the results presented in \citet{Carter1981}, who theoretically showed that return values estimates from sub-samples of the population are lower than the estimates from sampling of the population. This highlights that the inclusion of covariates improves the marginal estimation of return values.

\section{River flow extremes for the Harbourne}\label{sec:CaSt}

\subsection{Strategy and River Flow Data}\label{sec:LarSca}
The methods and risk measure developed in Sections \ref{sec:EVT} and \ref{sec:RMea} are applied to river flow data from the river Harbourne in a catchment in South Devon in the South West of England. As discussed in Section \ref{sec:Intro}, England's Environment Agency have focused on trying to reduce the risk of flooding to the communities situated along this river. They built a flood defence at Palmer's Dam, upstream of Harbertonford, to create extra capacity and reduce the risk of flooding \citep{EA2012}. However, Palmer's Dam was overtopped in 2012 by an event estimated to be a 1 in 40 year event, which caused further flooding to properties in the local area \citep{Devon2013}. 

Daily mean river flow data from 1998-2017 for the Rolster Bridge gauging station on the river Harbourne are considered. Exploratory analysis of these data can be found on the National River Flow Archive website \citep{NRFA}. The data were declustered above the $97$th percentile with an event window of $w$=7 days \citep{Keef2009} to produce on average 3 independent events per year. The methods described in Section \ref{sec:ObsCov} were used to model the declustered data, however there was no statistically significant evidence to include random effects in the parameters of the Poisson process possibly due to a relatively short record length of 20 years. However, this picture changes if we examine evidence for random effects by using data from neighbouring sites with longer record lengths. This longer record length and an assumption of a common random effect across a region allows us to pool information to obtain more reliable estimates of the random effects. The regional random effect represents a function of large scale weather events, which induces common shared behaviour between river flow gauges in a spatially homogeneous region. 

\begin{figure}[!htbp]
\centering
\subfloat[][]{\label{fig:Spa}
\includegraphics[width=7cm,height=6cm]{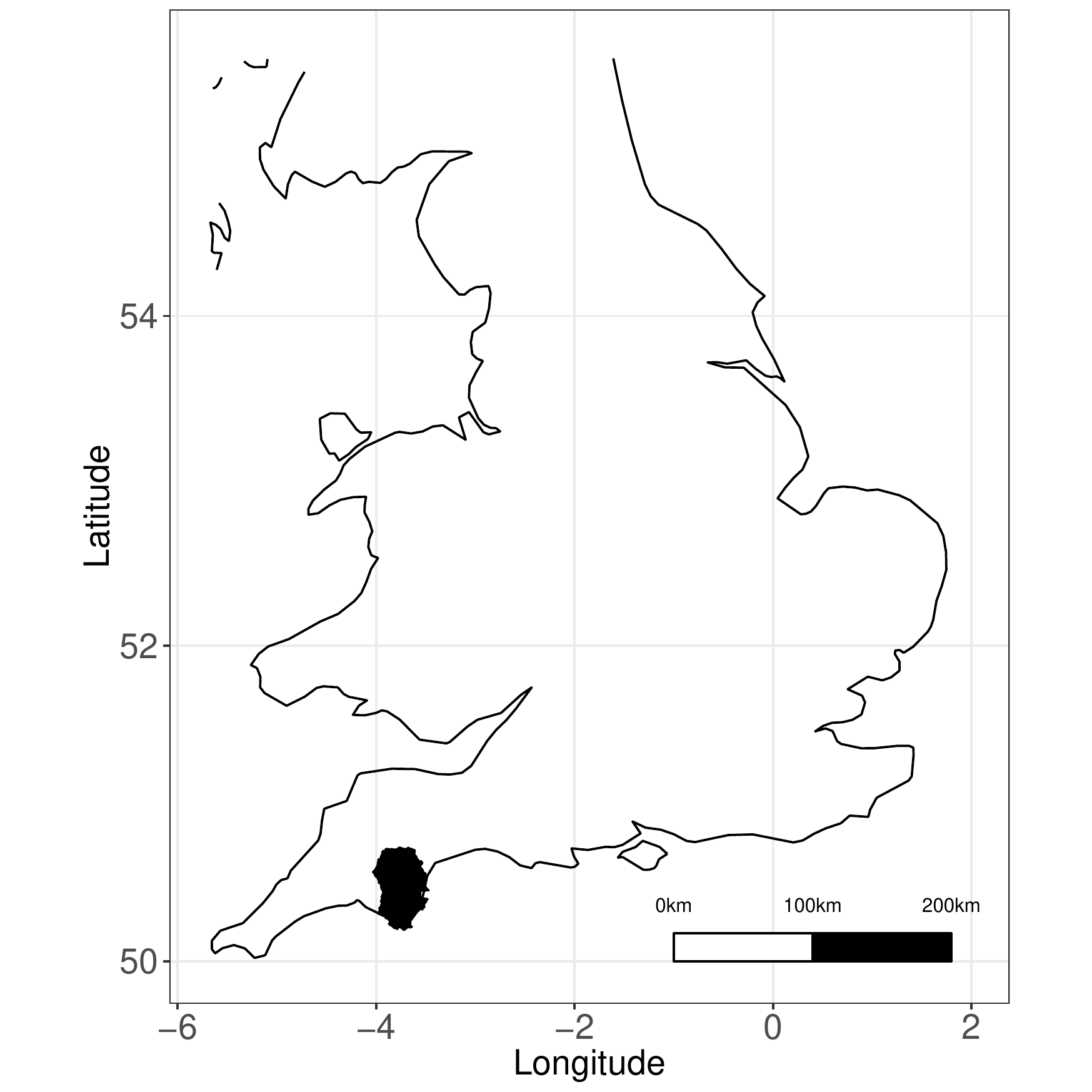}
}
\subfloat[][]{\label{fig:Spb}
\includegraphics[width=7cm,height=6cm]{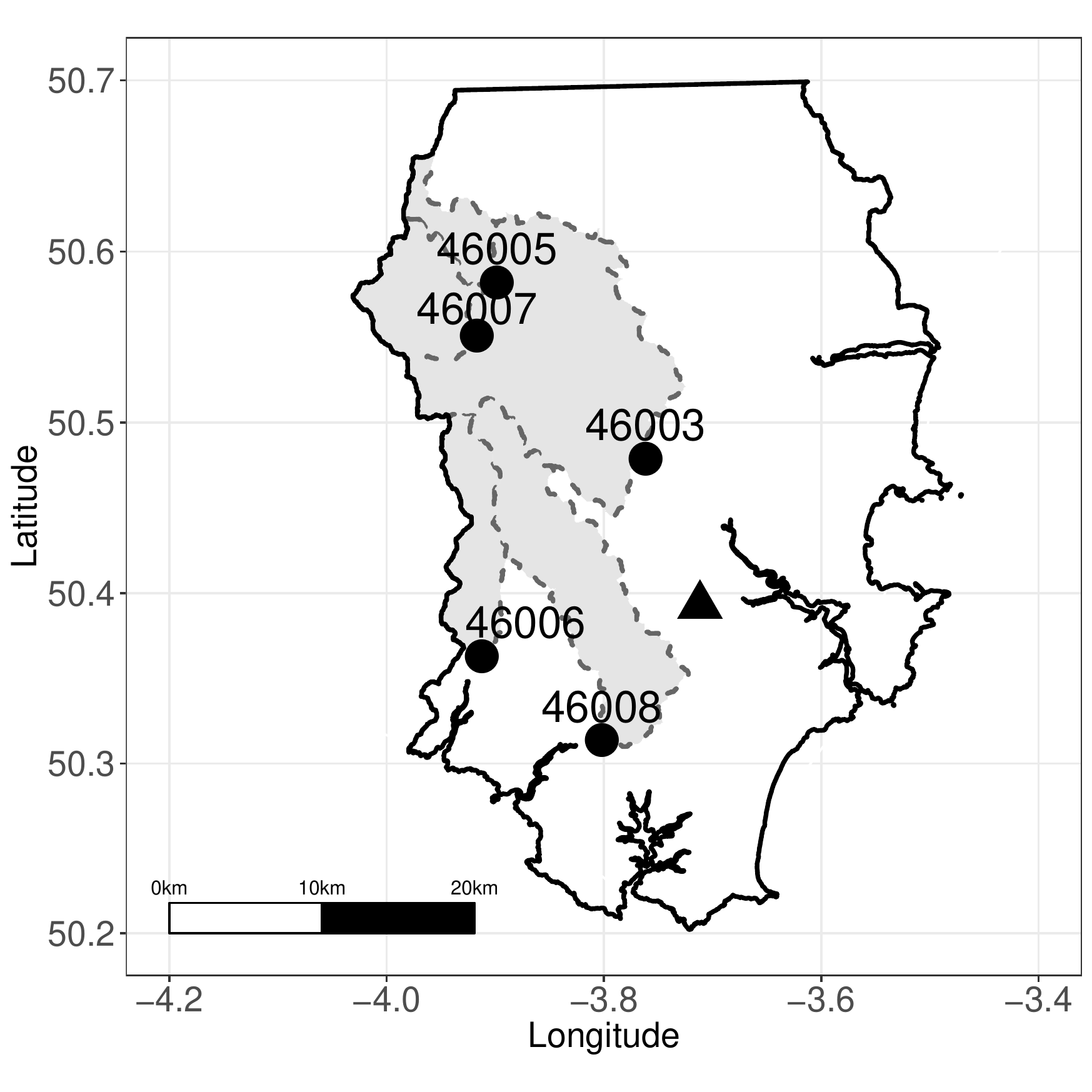}
}
\caption{Hydrometric region 46, (a): location within England and (b): region map. The triangle shows the location of the gauge on the river Harbourne and the circles show the locations of NRFA gauges. The grey shaded areas are the catchments of the NRFA gauges. Further details of the gauges can be found by searching their catalogue numbers (shown on the map) at nrfa.ceh.ac.uk.}
\label{fig:VisData}
\end{figure}

The additional data set consists of peak flows from five nearby river flow gauges, for the water years 1958-2013 (year defined from October to September) in the National River Flow Archive \citep{NRFA2014}. These gauges are all situated within the same hydrometric area as the river Harbourne, and so share similar physical characteristics, see Figure~\ref{fig:VisData}. However, they do not include any stations that are situated on the river Harbourne. These peak flow data sets are declustered to produce on average 5 independent events ever year. 


\subsection{Regional Random Effects Model}\label{sec:Mod}
The random effects model of Section \ref{sec:UnObsCov} can be adopted to include a common random effect across a number of locations, using the methods outlined in \citet{Eastoe2018} who develops an equivalent model for the generalised Pareto distribution. The model given in equation \eqref{eq:parset} is extended to include both site specific effects as well as a common regional random effect across sites. Consider a set of sites that are located in a spatially homogeneous region with NHPP parameters for site $d$ and year $i$ given by
\vspace{-0.8cm}
\begin{center}
\begin{eqnarray*}
\mu_{d}(r_{\mu,i})&=&\mu_{d,0}+\mu_{1}r_{\mu,i};~\log\left[\sigma_{d}(r_{\sigma,i})\right]=\sigma_{d,0}+\sigma_{1}r_{\sigma,i};~\xi_{d}(r_{\xi,i})=\xi_{d,0}+\xi_{1}r_{\xi,i}
\nonumber
\end{eqnarray*}
\end{center}
where for a given year $i$ the random effects $\mathbf{r}_{i}=(r_{\mu,i},r_{\sigma,i},r_{\xi,i})$ have the joint distribution defined in equation \eqref{eq:ReDist}. The parameters $\mu_{d,0}$, $\sigma_{d,0}$ and $\xi_{d,0}$ are site-specific intercept terms that account for differences between the gauging stations, e.g., the catchment size. The parameters $\mu_{1}$, $\sigma_{1}$ and $\xi_{1}$, along with the random effects, are common across sites. Under the assumption of spatial independence of observations at different sites in year $i$, given the random effects $\mathbf{r}_{i}$ for that year, coupled with independence of $\mathbf{r}_{1},\ldots,\mathbf{r}_{n_{y}}$, the likelihood function over sites is
\vspace{-1cm}
\begin{center}
\begin{eqnarray}
L(\boldsymbol{\mu},\boldsymbol{\sigma},\boldsymbol{\xi},\mathbf{r}_{1:n_{y}},\Sigma;\mathbf{z}_{1},\ldots,\mathbf{z}_{d})~\propto~\exp\left\{-\sum^{n_{d}}_{d=1}\sum^{n_{y}}_{i=1}\left[1+\xi_{d}(r_{\xi,i})\left(\frac{u_{d}-\mu_{d}(r_{\mu,i})}{\sigma_{d}(r_{\sigma,i})}\right)\right]^{-\frac{1}{\xi_{d}(r_{\xi,i})}}_{+}\right\} \nonumber\\
\times\left[\prod^{n_{y}}_{i=1}\left\{\prod^{n_{d}}_{d=1}\prod^{n_{u}(i,d)}_{j=1}~\frac{1}{\sigma_{d}(r_{\sigma,i})}\left[1+\xi_{d}(r_{\xi,i})\left(\frac{z_{dij}-\mu_{d}(r_{\mu,i})}{\sigma_{d}(r_{\sigma,i})}\right)\right]^{-\frac{1}{\xi_{d}(r_{\xi,i})}-1}_{+}\right\}\phi(r_{\mu,i},r_{\sigma,i},r_{\xi,i};\Sigma)\right],
\label{eq:ReLRe}
\end{eqnarray}
\end{center}

where $n_{u}(i,d)$ is the number of exceedances of threshold $u_{d}$ in year $i$ at site $d$ and $\mathbf{z}_{j}$ are these $\sum^{n_{y}}_{i=1}n_{u}(i,d)$ exceedances and $z_{dij}$ is the $j$th of $n_{u}(i,d)$ exceedances of $u_{d}$ at site $d$ in year $i$. 

Model \eqref{eq:ReLRe} is fitted under a Bayesian framework to the NRFA peak river flow data of Section \ref{sec:LarSca}, using an adaptive MCMC algorithm which was run for 200,000 iterations with a burn-in of 50,000. Uninformative priors are used for all parameters of the Poisson process. It was found that there was no evidence to incorporate random effects into the shape parameter, therefore random effects were only incorporated into the location and scale parameters. This is a typical finding with covariates rarely found to be statistically significant for the shape parameter in environmental applications \citep{Eastoe2009,Eastoe2018}.  As a result of this the random effects model of Section \ref{sec:UnObsCov} is simplified so that $\mathbf{r}_{i}=(r_{\mu,i},r_{\sigma,i})$ follows a standard bivariate Normal with correlation parameter $\rho$ and $\xi_{d}$ is the shape parameter for site $d$. 

\begin{figure}[!h]
\centering
\vspace{-0.2cm}
\subfloat[][]{\label{fig:muRE}
\includegraphics[width=8cm,height=7cm]{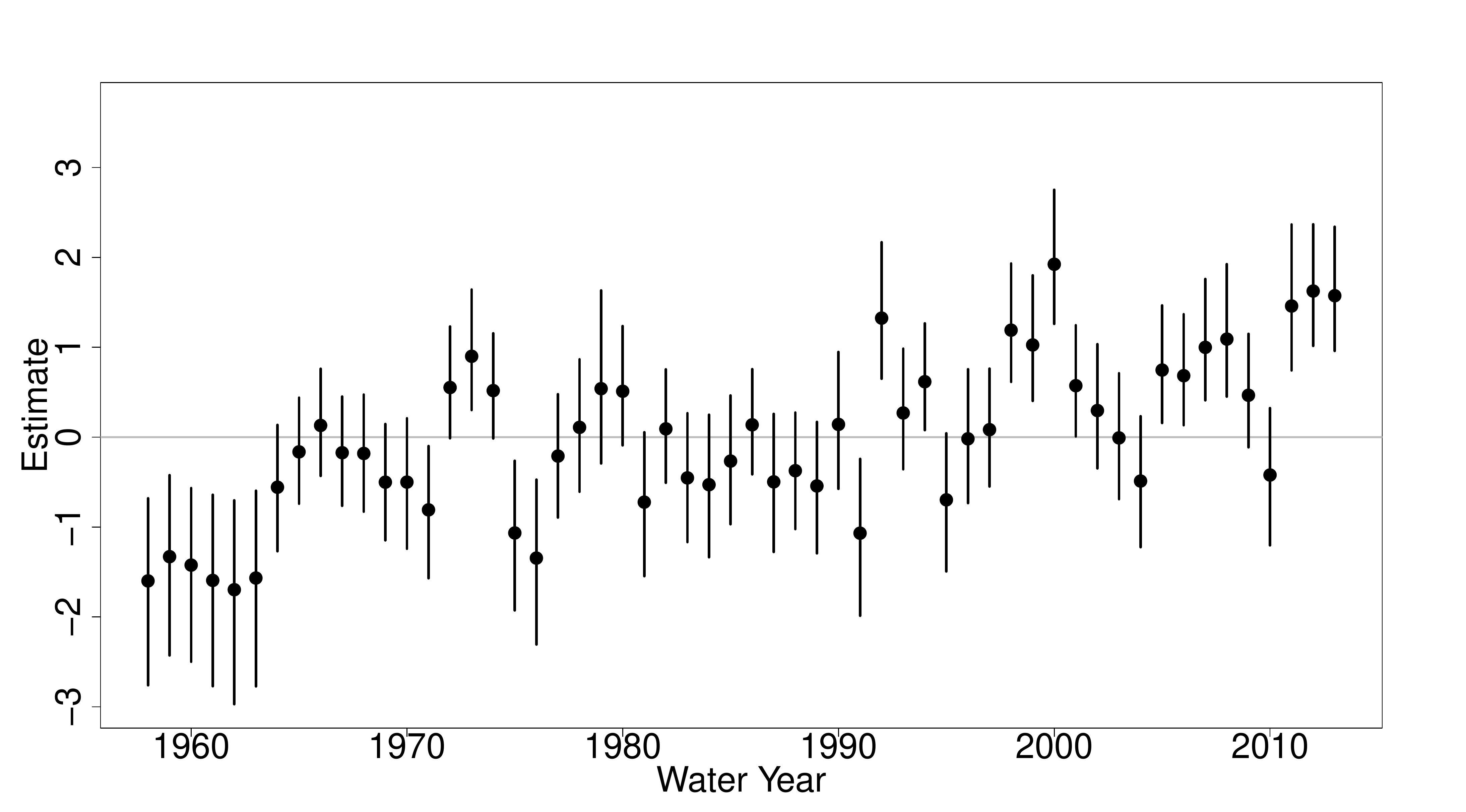}
}
\subfloat[][]{\label{fig:sigRE}
\includegraphics[width=8cm,height=7cm]{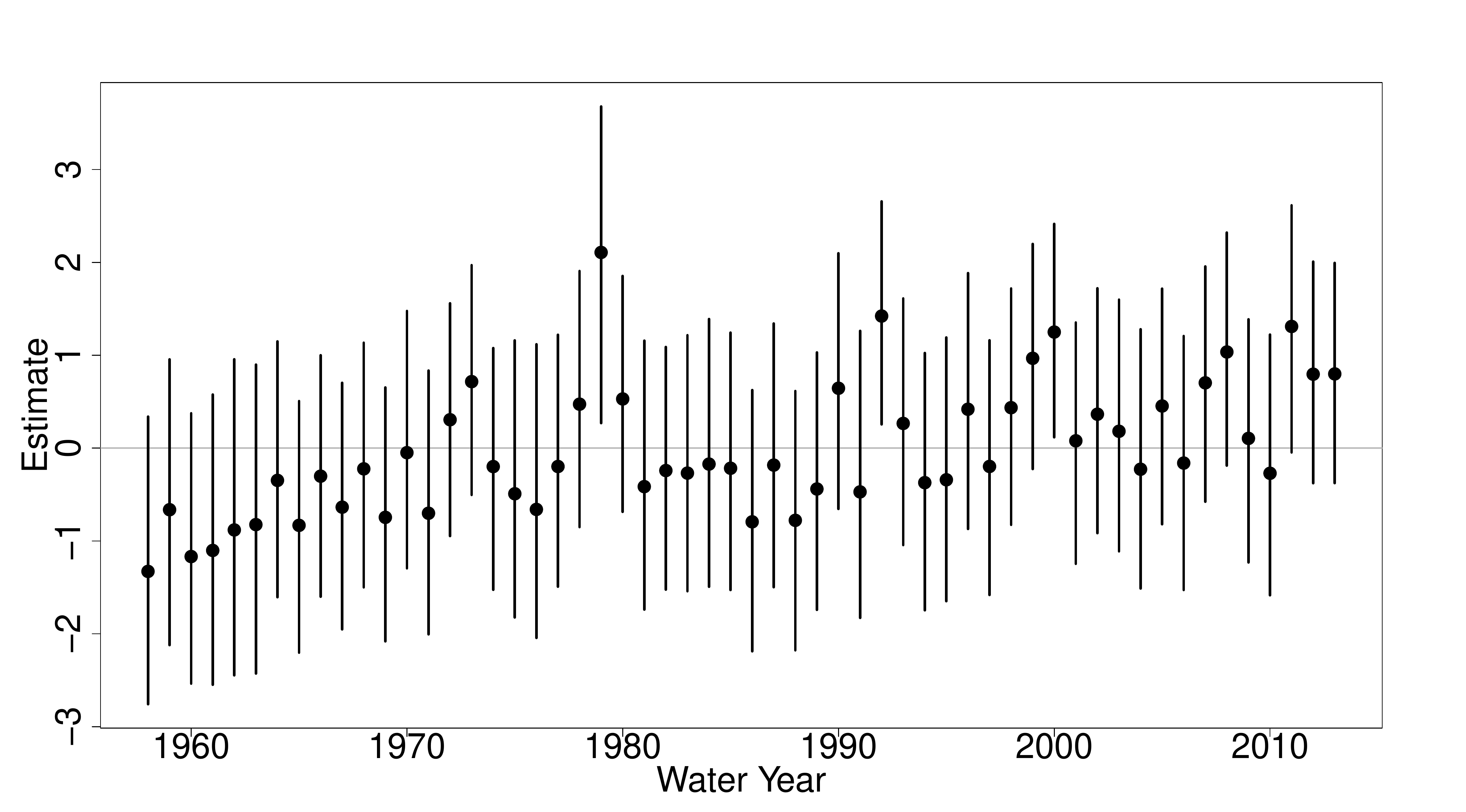}
}\\
\caption{Random effect estimates for the location (a) and scale (b) parameters plotted against the respective water years. The vertical lines represent 95$\%$ credible intervals.}
\label{fig:RE}
\end{figure}

The estimates in Figures \ref{fig:muRE} and \ref{fig:sigRE} show that it is unsuitable to assume that the location and scale parameter of the Poisson process remain constant over years. In particular the largest values for the location random effect correspond to known flood events such as those in the water years 1990 and 2003 \citep{Devon2013}. The location parameter shows much greater variability across years than the scale parameter. For the random effect in the scale parameter there is a clear positive value for 1979, which corresponds to known floods in the region \citep{Devon2013}. In comparison, there is no significant location random effect estimate for 1979 as the credible interval contains the value zero. This particular year has the largest flow across all river flow gauges and a particularly large value of NAO, which corresponds to wet winters. There are more typical size extreme events as well as this large event and the large value of the random effect in the scale parameter captures both of these features. A similar change in the random effect of the location parameter would be insufficient as it could not cover the associated change in the variance of the threshold excesses in this year. The correlation of the random effects was estimated to be 0.62 with a 95$\%$ credible interval of $(0.13,0.88)$. The autocorrelation of both sequences of random effects was explored to see whether we could extend the model by using an autoregressive process to model the random effects. However, there was no evidence that any serial correlation exists and so we retain the simpler model in which the random effects are assumed to be independent over years. There is also no significant relationship between the random effects and the climate index NAO.  


\subsection{Short-term Risk Measure Estimates}\label{sec:ReRa}
\begin{figure}[!htbp]
\centering
\vspace{-0.2cm}
\subfloat[]{\label{fig:r1N}
\includegraphics[width=5cm,height=5cm]{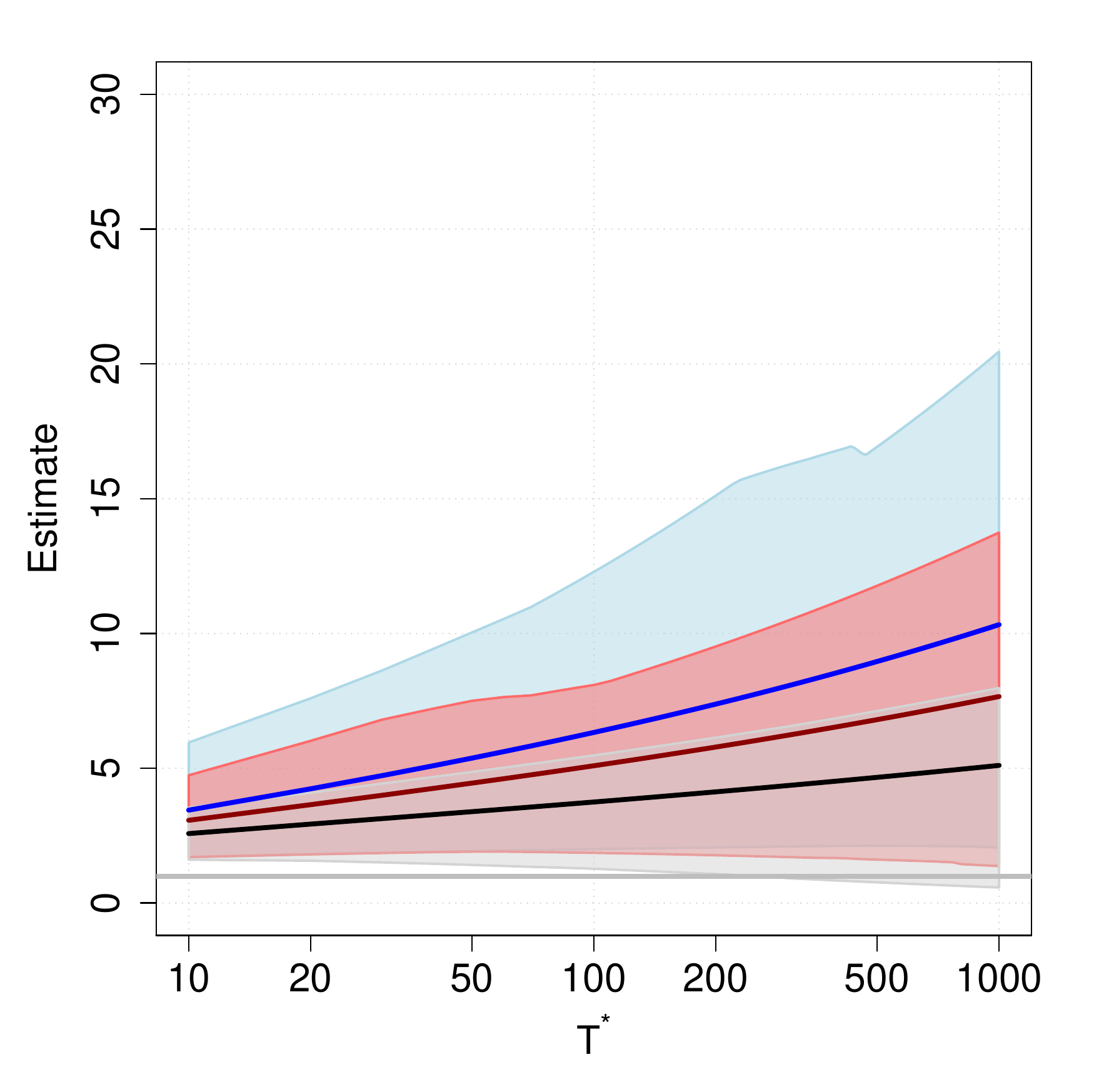}
}
\subfloat[]{\label{fig:r2N}
\includegraphics[width=5cm,height=5cm]{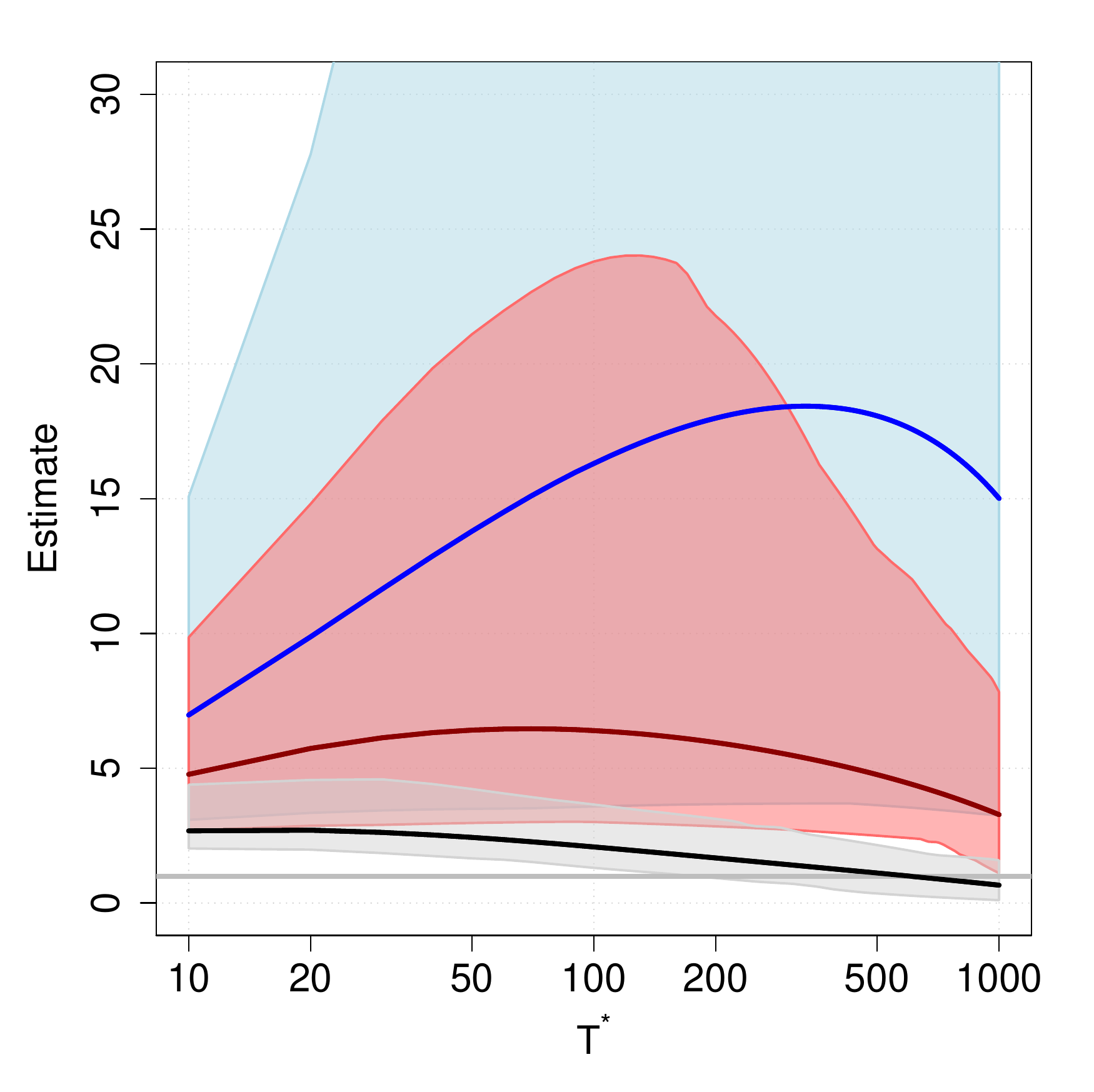}
}\\
\subfloat[]{\label{fig:r1A}
\includegraphics[width=5cm,height=5cm]{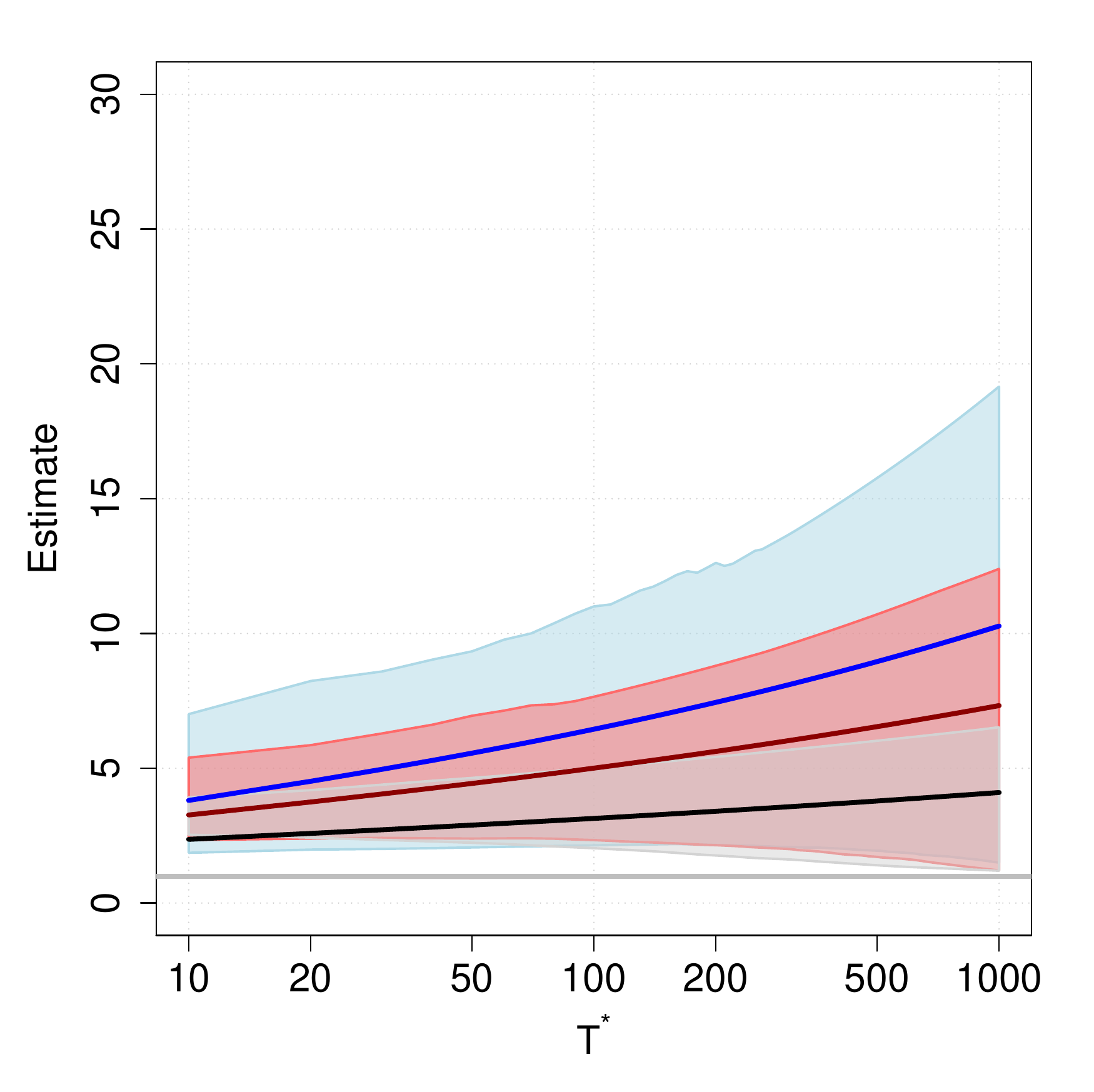}
}
\subfloat[]{\label{fig:r1A}
\includegraphics[width=5cm,height=5cm]{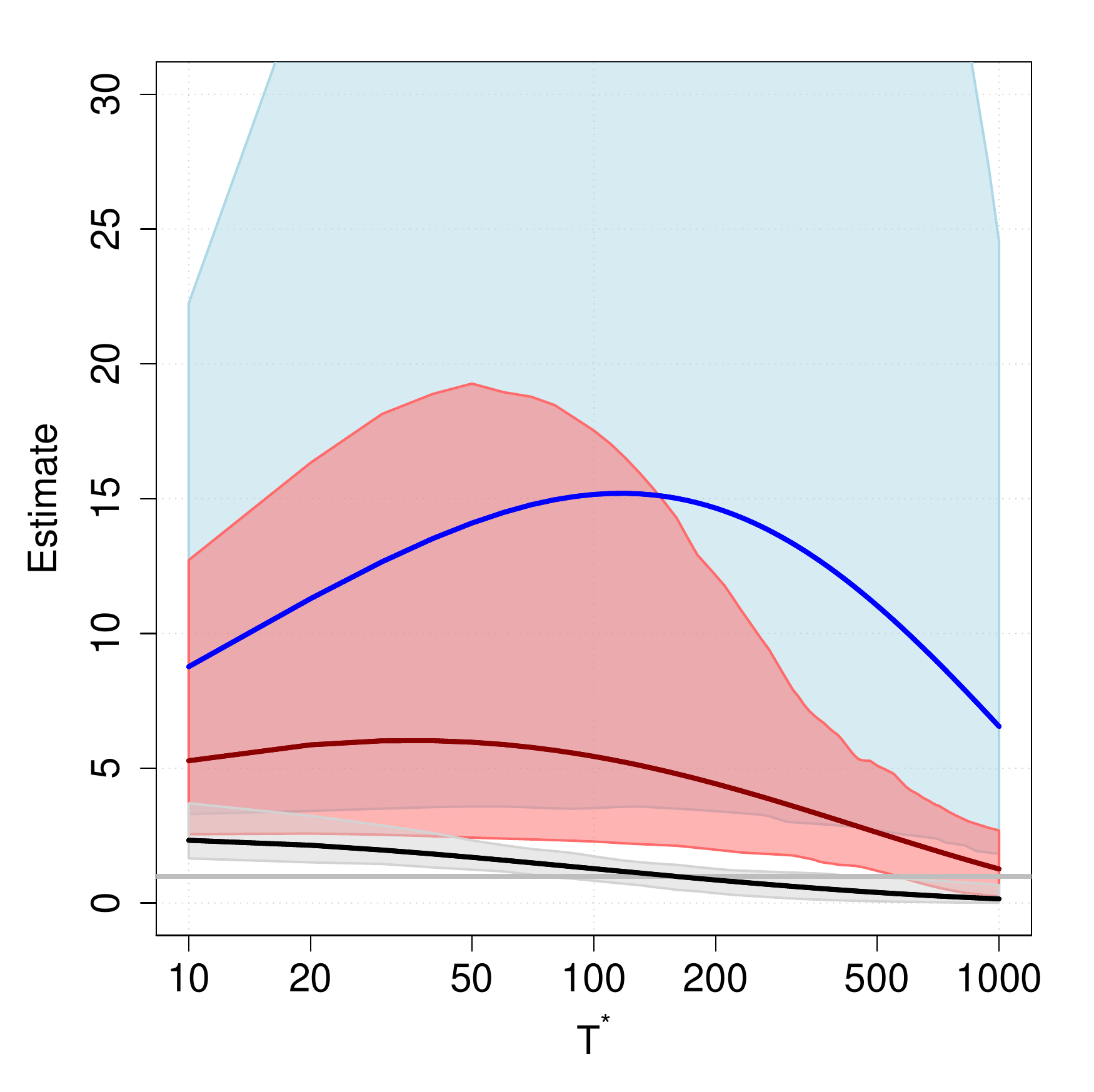}
}
\caption{Short-term risk measure $R^{(t)}_{T,T^{*}}$ for gauges 46006 (a,c) and 46008 (b,d) if an extreme event occurred in December $(t=0.20)$ or June $(t=0.80)$ (top and bottom row respectively). The black, red and blue lines correspond to already observing a 1 in $T$=10, 100 and 1000 year event with the corresponding shaded areas providing 95$\%$ credible intervals.}
\label{fig:RRNov}
\end{figure}

We now estimate the short-term risk measure derived in Section \ref{sec:RMea} for the two of the five peak flow river gauges discussed in Section \ref{sec:Mod}. Figure \ref{fig:RRNov} presents how $R^{(t)}_{T,T^{*}}$ changes depending on $t$, $T$ and $T^{*}$ for each gauge. Two different points, December and June, are considered for the time $t$ with $T=10,100,1000$. The short-term risk measure estimates are larger when conditioning on the more severe events. For each gauge, there is a significant increase in risk for the 100 and 1000 year events as shown in Figure \ref{fig:RRNov}. We can see the trade-off between the size (return period $T$) and timing, $t$, of the event in influencing the chance of observing extreme event of return period $T^{*}$, with $R^{(t)}_{T,T^{*}}$ appearing to decrease as a function of $t$. As expected the confidence interval is widest when $T=1000$ years. If we now condition on an extreme event occurring later in the water year, in this case June, the short-term risk measure has decreased due to there being a smaller amount of time remaining in the season to observe another extreme event. 
Although the regional random effects are common across the gauges, the risk estimates are different due to the site-specific parameters of the non-homogeneous Poisson process: so the risk measure captures the different effects of extreme events at each site. 

\subsection{Relating the results to the River Harbourne}\label{sec:RiHar}
\begin{figure}[!htbp]
\centering
\subfloat[]{\label{fig:mu2}
\includegraphics[width=5cm,height=5cm]{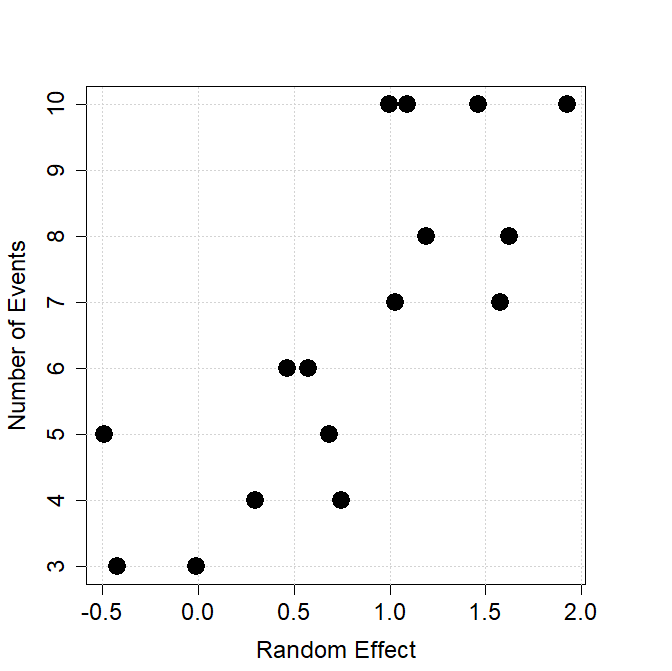}
}
\subfloat[]{\label{fig:sig2}
\includegraphics[width=5cm,height=5cm]{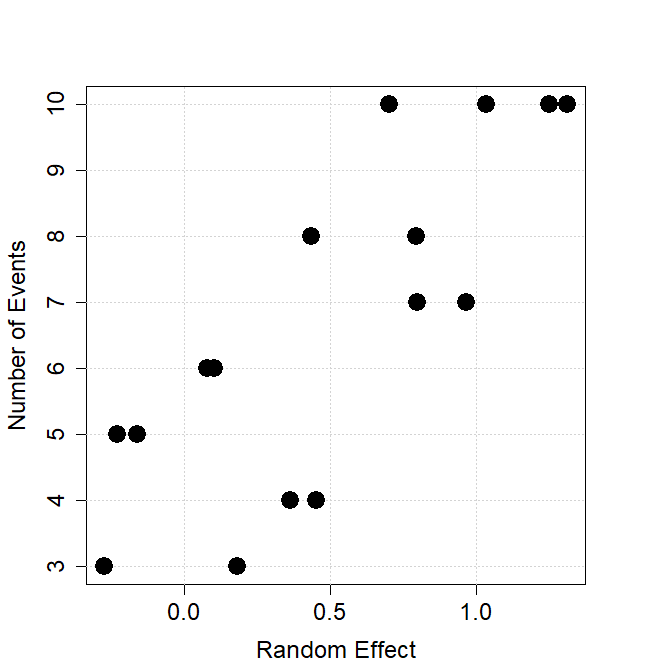}
}\\
\caption{Comparison of the regional random effects estimate for the (a) location and (b) scale parameters against the number of events for the Rolster Bridge gauging station from the river Harbourne.}
\label{fig:RolsterRE}
\end{figure}

To assess the value of the annual regional random effect estimates $(r_{\mu,i},r_{\sigma,i})$, we compare them with the associated annual number of threshold exceedances for the river Harbourne in Figure \ref{fig:RolsterRE}. For each random effect there does seem to be an association, which is stronger for the location parameter. 

In order to assess the validity of the random effects shown in Figure \ref{fig:RE}, the random effect estimates are taken as fixed and the model given in equation \eqref{eq:parset} is estimated for the gauge on the river Harbourne, denoted by site $H$. The Poisson process parameter estimates of $\mu_{1,H}$ and $\sigma_{1,H}$ are significantly different from zero, suggesting we have good covariates in the form of $(r_{\mu,i},r_{\sigma,i})$. However, the estimate of the local location parameter $\mu_{1,H}$ for the spatial random effect was statistically significantly less than $\mu_{1}$ (the regional location parameter estimate for the spatial random effect) so not all aspects of the regional model translate to the river Harbourne. 
\vspace{-0.5cm}

\begin{figure}[!htbp]
\centering
\vspace{-0.2cm}
\subfloat[]{\label{fig:rHN}
\includegraphics[width=6.5cm,height=6.5cm]{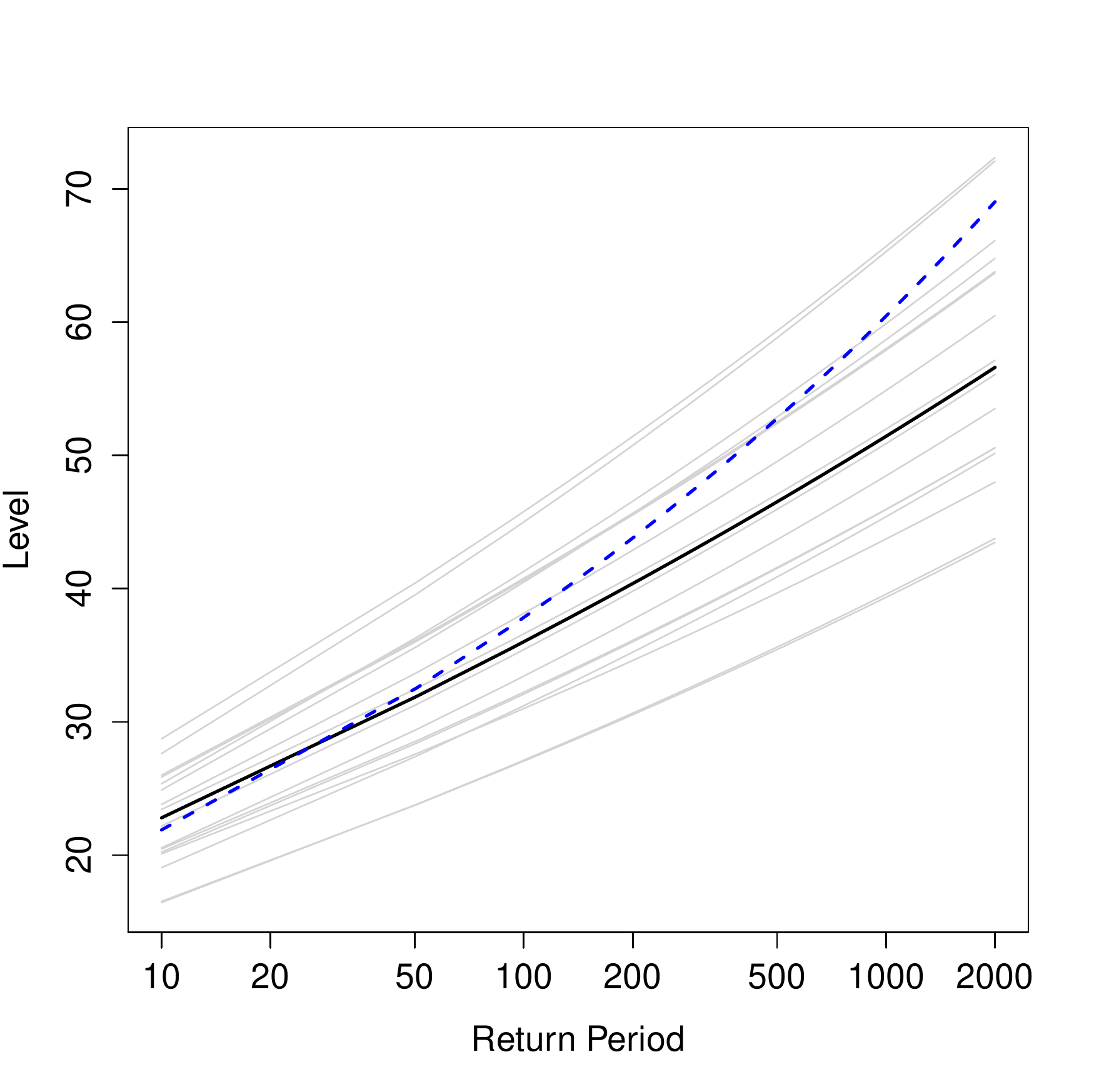}
}
\subfloat[]{\label{fig:rHJ}
\includegraphics[width=6cm,height=6cm]{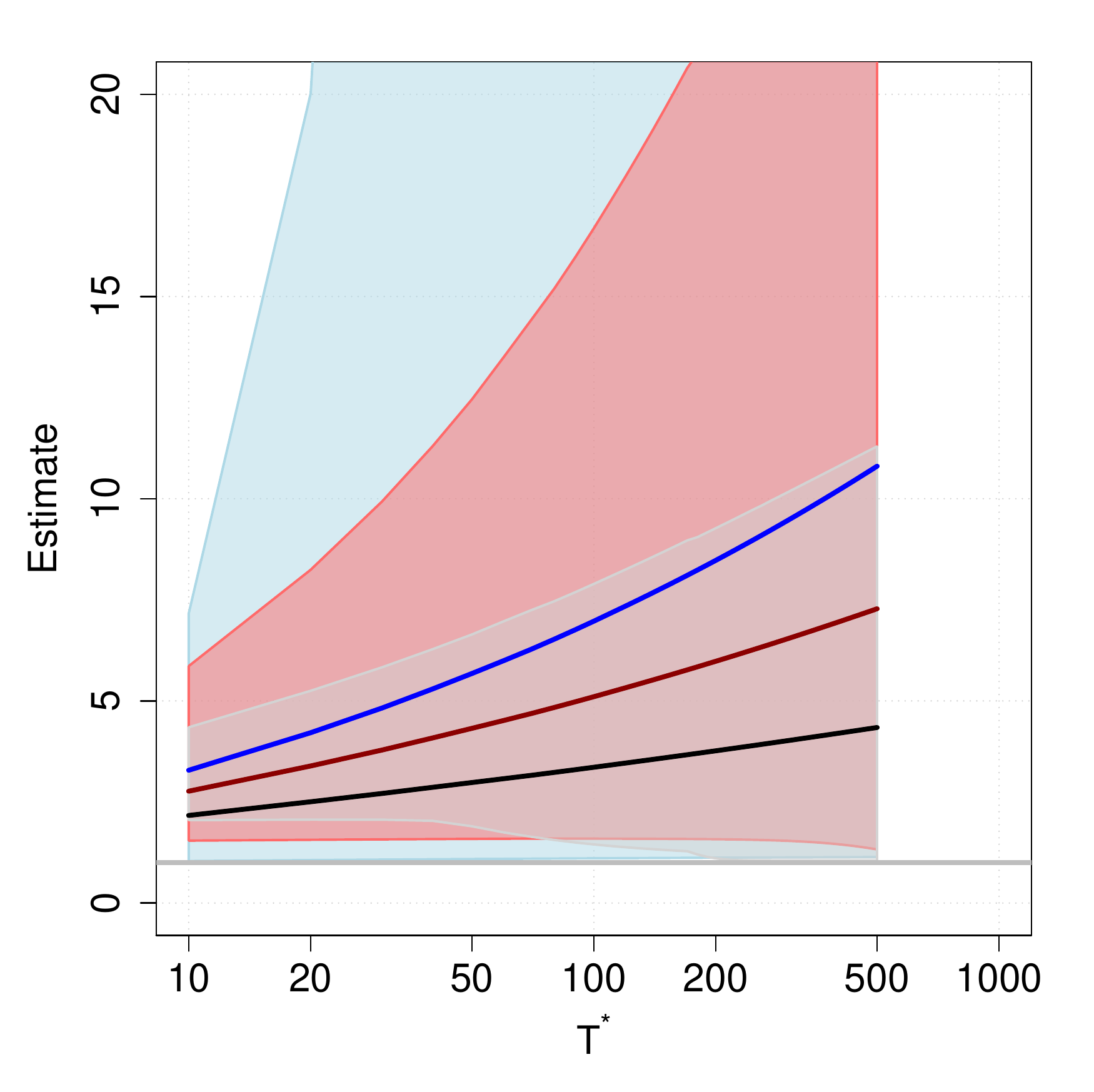}
}
\vspace{-0.2cm}
\caption{(a): return level estimates for the gauge on the river Harbourne. The blue dashed line is the estimate from fitting a GEV distribution to the annual maxima. The grey lines show the return level estimates for each year conditional on its random effect estimate with the black line showing the cross-year posterior averages; (b): short-term risk measure $R^{(t)}_{T,T^{*}}$ for the river Harbourne if an extreme event occurred in December $(t=0.20)$. The black, red and blue lines correspond to already observing a 1 in $T$=10, 100 and 1000 year event with the corresponding shaded areas providing 95$\%$ credible intervals.}
\label{fig:rtHarb}
\end{figure}

Figure \ref{fig:rtHarb} shows both return level and risk measure plots for the Harbourne derived from using the regional random efforts and treating these as fixed. Figure \ref{fig:rtHarb}a shows the return level curve for each annual random effect, illustrating how much the annual maximum distribution varies over time, as well as the time averaged estimate which gives us a best estimate for the future return levels when excluding long-term climate changes. For comparison we also show the estimated return level curve based on a GEV fit to the annual maxima from the Harbourne, similarly to Figure~\ref{fig:DisMax}. Again we find that relative to fitting with random effects this GEV fit overestimates the risk of large events, i.e., events with return periods of more than 100 years. The risk measure estimates, shown in Figure \ref{fig:rtHarb}b, show that the short-term risk of flooding for the Harbourne increases as the size of the conditioning event increases. For example, for a risk assessment in December ($20\%$ of the way into the storm season) the increased risk of a 100-year event ($T^{*}=100$) in the remainder of the storm season doubles from 3 to 6 if the previously observed event had a return period of $T=10$ or 1000 years, i.e., these updated odds of these events for the rest of the storm season are $1:33$ to $1:17$. 

\vspace{-1cm}

\pagebreak

\section{Discussion}\label{sec:Dis}
There has been much coverage in the media of the reoccurrence of extreme events at a single location. The common method of communicating the severity of these events is the return period. However, this definition is only valid when the process is stationary. The clustering of independent extreme events in a short period of time provides clear evidence that the assumption of a homogeneous Poisson process for the occurrence of extreme events is no longer valid. As a result, using return periods to communicate the severity of an event is no longer suitable for short-term risk assessment. 

We have developed a short-term risk measure to help convey the change in the risk of observing extreme events later in a storm season after observing an extreme event. This measure captures the local non-stationarity in time within the data. This local non-stationarity can be explained through incorporating covariate information, either from observations or through the use of random effects. The covariate information allows us to improve the accuracy of the statistical models and ultimately produce more reliable and informative return value estimates. This novel methodology helps to provide an advancement towards improving the long-term modelling of extreme flooding in the presence of time variability, an issue raised by the National Flood Resilience Review \citep{NFRR2016, Tawn2018, Towe2018}. 

The methodology was applied to river flow data from South Devon. The models that are fitted in Section \ref{sec:CaSt} show that there is clear evidence of a similar inter-year non-stationarity over sites in this area. These random effects estimates were then used to estimate the short-term risk measure. For a hydrometric area we have shown that there is a clear change in risk of observing an extreme event once one has already been observed. Naturally, the magnitude of this change in risk is a function of the size of the original event and the time at which it was observed. 


The short-term risk measure developed in Section~\ref{sec:ViRe} is one particular method of assessing risk but alternative risk measures could also be considered. We may be interested in the probability of there being an exceedance of $z_{T}$ for the remainder of the time period given that there have been $n$ exceedances of $z_{T}$ so far in the year. Figure \ref{fig:AltRM} illustrates this event with $n=3$. The risk measure could then be constructed as the ratio of this probability with the marginal probability $\mathbb{P}(M_{\lfloor tn \rfloor :n}>z_{T})$.


\renewcommand{\abstractname}{Acknowledgements}
\begin{abstract}
Towe's research was supported by the NERC PURE Associates program and EP/P002285/1 (The Role of Digital Technology in Understanding, Mitigating and Adapting to Environmental Change). We would also like to thank Ilaria Prosdocimi (University of Bath) for providing and pre-processing the National River Flow Archive peak flow data. The data for the river Harbourne were provided by Tim Shipton from the Environment Agency. We are grateful for help from Maxine Zaidman in identifying this case study and referees for their recommendations to improve the presentation. 
\end{abstract}

\bibliographystyle{apalike}			
\bibliography{References}				
\end{document}